\documentclass[%
 showpacs,
 showkeys,
 amsmath,amssymb,
]{revtex4-1}

\usepackage{graphicx}
\usepackage{dcolumn}
\usepackage{bm}

\usepackage{CJK}

\usepackage{color}
\usepackage{amssymb}
\usepackage{amsfonts}

\usepackage[colorlinks,urlcolor=blue,linkcolor=blue,citecolor=blue,anchorcolor=blue]{hyperref}

\begin{document}


\preprint{APS/123-QED}

\title{Controllable circular Airy beams via dynamic linear potential}

\author{Hua Zhong$^1$}
\author{Yiqi Zhang$^1$}
\email{zhangyiqi@mail.xjtu.edu.cn}
\author{Milivoj R. Beli\'c$^{2}$}
\author{Changbiao Li$^1$}
\author{Feng Wen$^1$}
\author{Zhaoyang Zhang$^1$}
\author{Yanpeng Zhang$^{1}$}
\affiliation{%
 $^1$Key Laboratory for Physical Electronics and Devices of the Ministry of Education \& Shaanxi Key Lab of Information Photonic Technique,
Xi'an Jiaotong University, Xi'an 710049, China \\
$^2$Science Program, Texas A\&M University at Qatar, P.O. Box 23874 Doha, Qatar
}%

\date{\today}

\begin{abstract}
  \noindent
  We investigate controllable spatial modulation of circular autofocusing Airy beams,
  under action of different dynamic linear potentials, both theoretically and numerically.
  We introduce a novel treatment method in which the circular Airy beam is represented as a superposition of narrow azimuthally-modulated one-dimensional Airy beams that can be analytically treated.
  The dynamic linear potentials are appropriately designed, so that
  the autofocusing effect can either be weakened or even eliminated when the linear potential exerts a ``pulling'' effect on the beam,
  or if the linear potential exerts a ``pushing'' effect, the autofocusing effect can  be greatly strengthened.
  Numerical simulations agree with the theoretical results very well.
\end{abstract}

\maketitle

%
\section{Introduction}

In quantum mechanics, the Airy function has been shown to be a solution of the potential-free Schr\"odinger equation \cite{berry.ajp.47.264.1979}.
Since the Airy wave function is not a realistic physical quantity, due to its infinite energy,
the investigation of Airy wavepackets in physics did not attract much attention until
the introduction of finite-energy Airy beams in optics \cite{siviloglou.ol.32.979.2007,siviloglou.prl.99.213901.2007}.
Such beams possess interesting and useful self-accelerating, self-healing \cite{broky.oe.16.12880.2008} and nondiffracting properties.
Hence, in the past decade topics related with Airy beams received intense attention.
Thus far, investigations of Airy beams have been reported in nonlinear media \cite{kaminer.prl.106.213903.2011,lotti.pra.84.021807.2011,dolev.prl.108.113903.2012,zhang.ol.38.4585.2013,zhang.oe.22.7160.2014,shen.sr.5.9814.2015,diebel.oe.23.24351.2015},
Bose-Einstein condensates \cite{efremidis.pra.87.043637.2013},
on the surface of a metal \cite{salandrino.ol.35.2082.2010,zhang.ol.36.3191.2011,minovich.prl.107.116802.2011,li.prl.107.126804.2011},
optical fibers \cite{hu.ol.38.380.2013,driben.ol.38.2499.2013,zhang.oe.23.2566.2015,hu.prl.114.073901.2015}, and other systems.
To manipulate the propagation of Airy beams, external potentials such as harmonic \cite{bandres.oe.15.16719.2007,zhang.oe.23.10467.2015,zhang.aop.363.305.2015}
or a linear potenial \cite{liu.ol.36.1164.2011,efremidis.ol.36.3006.2011}, have been introduced.
For more details, the reader may consult review articles \cite{hu.book.2012,zhang.csb.58.3513.2013,bandres.opn.24.30.2013} and references therein.

On the other hand, in more than one dimension the circular Airy (CAi) beams, as radially symmetric beams that possess autofocusing property \cite{efremidis.ol.35.4045.2010,papazoglou.ol.36.1842.2011,chremmos.ol.36.1890.2011,penciu.ol.41.1042.2016},
have stirred wide interest in the past few years.
It has been demonstrated that CAi beams can be used to trap and guide microparticles \cite{zhang.ol.36.2883.2011,li.oe.22.7598.2014}
and even to produce the so-called bottle beams \cite{chremmos.ol.36.3675.2011} and light bullets \cite{Panagiotopoulos.nc.4.2622.2013}.
Similar to the finite-energy Airy beams, CAi beams can also be controlled and manipulated by potentials \cite{hwang.pj.4.174.2012,ioannis.pra.85.023828.2012,zhang.ol.40.3786.2015}.
Recent research has shown that the propagation trajectory as well as the position of autofocusing points of the CAi beam can be controlled by a dynamic linear potential \cite{hwang.pj.4.174.2012}.
However, for fuller understanding further investigation of CAi beams is required. Interesting questions remain unanswered.
Can the layers of CAi beams be modulated during propagation?
Can the autofocusing effect be weakened or maybe strengthened?
We answer these questions in this article.

Here, we introduce specific linear potentials and investigate the corresponding influence on the propagation of CAi beams.
Since the linear potential is similar to a ``force'' in the classical mechanics \cite{berry.ajp.47.264.1979},
it will pull or push the CAi beams during propagation.
We give an analytical solution for a two-dimensional CAi beam propagating in the linear potential by an analogy to that for a one-dimensional finite-energy Airy beam,
and then carry out numerical simulations to verify its validity.
We find that the autofocusing can be strengthened when the linear potential plays a ``pushing'' role,
while when it plays a ``pulling'' role, the autofocusing effect can be weakened or even eliminated.

The setup of the article is as follows.
In Sec. \ref{theory}, we introduce the theoretical model,
and in Sec. \ref{results}, we present our theoretical and numerical results for different linear potentials.
In Sec. \ref{gauss}, different from Sec. \ref{results} where an exponential apodization is adopted,
we briefly discuss the case with a Gaussian apodization.
We conclude the paper and display some extended discussions in Sec. \ref{conclusion}.

\section{Theoretical model}\label{theory}
\begin{figure}[htbp]
\centering
  \includegraphics[width=0.7\columnwidth]{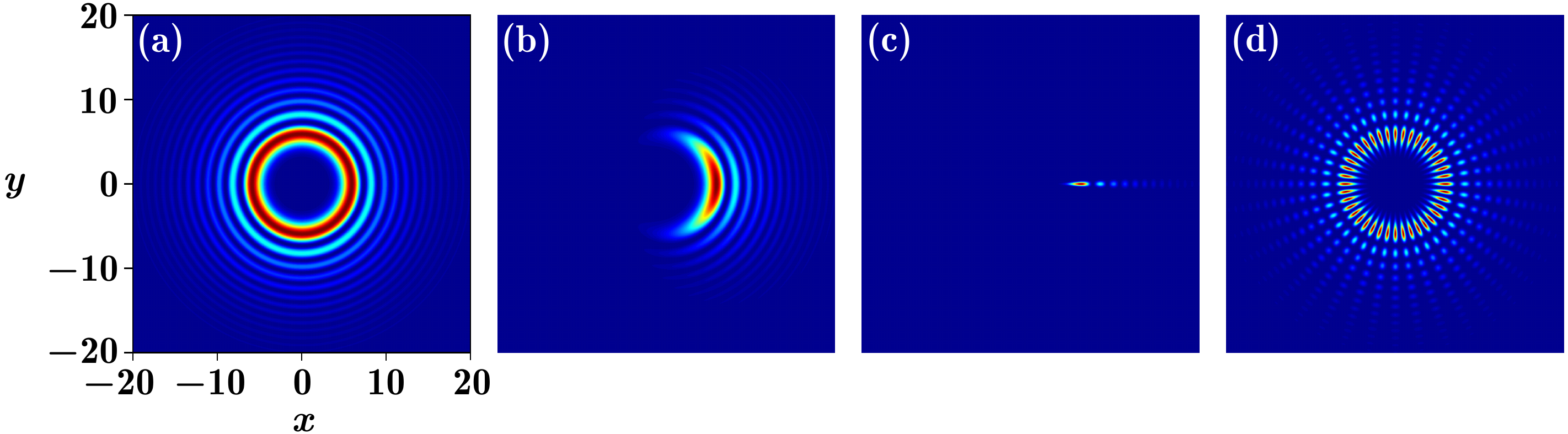}
  \caption{Input beam intensities.
  (a) An example of finite-energy circular Airy beam.
  (b) An azimuthally-modulated circular Airy beam, with $w_0=1.5$.
  (c) Same as (b), but with $w_0=0.05$.
  (d) A combination of 40 azimuthally modulated circular Airy beams along different azimuthal directions, according to Eq. (\ref{eq6}).
  Other parameters: $r_0=5$ and $a=0.1$. All the panels share the same variables and dimensions.
  }
  \label{fig1}
\end{figure}

The propagation dynamics of the CAi beam is described by the following Schr\"odinger equation:
\begin{equation}\label{eq1}
i\frac{d\psi }{dz}+\frac{1}{2}\left( \frac{{{\partial }^{2}}\psi }{\partial {{x}^{2}}}+\frac{{{\partial }^{2}}\psi }{\partial {{y}^{2}}} \right)-V(x,y,z)\psi =0,
\end{equation}
where $\psi$ is the beam envelope, $x$ and $y$ are the normalized transverse coordinates, and $z$ is the propagation distance,
scaled by some characteristic transverse width $x_0$ and the corresponding Rayleigh length $kx_0^2$, respectively.
Here, $k=2\pi n/\lambda_0$ is the wavenumber, $n$ the index of refraction, and $\lambda_0$ the wavelength in free space.
For our purposes, the values of parameters can be taken as $x_0=100\,\mu \rm m$, and $\lambda_0=600\,\rm nm$.
The external potential is chosen in a specific form $V(x,y,z)=d(z)r/2$, where $r=\sqrt{x^2+y^2}$. Thus, it is linear in $r$, with the scaling factor $d$ depending on the longitudinal coordinate.
In this form, it is known as the {\it dynamical} linear potential \cite{efremidis.ol.36.3006.2011,hwang.pj.4.174.2012,ioannis.pra.85.023828.2012}.
In polar coordinates, Eq. (\ref{eq1}) can be rewritten as
\begin{equation}\label{eq2}
i\frac{d\psi }{dz}+\frac{1}{2}\left( \frac{{{\partial }^{2}}\psi }{\partial {{r}^{2}}} + \frac{1}{r} \frac{\partial \psi }{\partial r}\right)-\frac{d(z)}{2}r \psi =0.
\end{equation}
In the (1+1)-dimensional limit, Eq. (\ref{eq1}) can be written as
\begin{equation}\label{eq3}
i\frac{\partial\psi}{\partial z}+\frac{1}{2}{\frac{\partial^2\psi}{\partial x^2}}-\frac{d(z)}{2}|x| \psi=0,
\end{equation}
which clearly is a linear potential. Without the absolute value sign on {$x$},  Eq. (\ref{eq3}) is the same as that treated in \cite{efremidis.ol.36.3006.2011}.
However, with the absolute value, one can manage this problem for $x>0$ and $x<0$ separately \cite{zhang.prl.115.180403.2015}.
In Eq. (\ref{eq1}) or Eq. (\ref{eq2}), we assume that the input beam is an inward CAi beam
\begin{equation}\label{eq4}
\psi(x,y)={\rm Ai}(r_0-r)\exp[a(r_0-r)],
\end{equation}
where $a$ is the decay parameter -- a positive real number that makes the total energy finite,
and ${{r}_{0}}$ is the initial radius of the main Airy ring.
For our purpose, we assume ${{r}_{0}}=5$; the corresponding intensity distribution of the input is shown in Fig. \ref{fig1}(a).
We note that it is hard to find an analytical solution of  Eq. (\ref{eq1}) or Eq. (\ref{eq2}) by the method developed in \cite{efremidis.ol.36.3006.2011},
because of the difficulty with the Fourier transform of the last term in Eq. (\ref{eq1}) or the zero-order Hankel transform of the last term in Eq. (\ref{eq2}).
Here, we develop an approximate but accurate method to solve this problem -- by introducing an azimuthal modulation of the CAi beam in Eq. (\ref{eq4}),
\begin{equation}\label{eq5}
{{\psi }_{\text{az}}}(x,y)=\text{Ai}({{r}_{0}}-r)\exp [a({{r}_{0}}-r)]\exp \left( -\frac{{{(\theta -{{\theta }_{0}})}^{2}}}{w_{0}^{2}} \right),
\end{equation}
with ${{w}_{0}}$ being the width of the modulation, $\theta =\arctan (y/x)$ being the azimuthal angle,
and $\theta_0$ indicating the modulation direction.

In Fig. \ref{fig1}, we depict some input possibilities.
First, we set ${{w}_{0}}=1.5$ and $\theta_0=0$;
the corresponding intensity is displayed in Fig. \ref{fig1}(b), which is a crescent-like structure with many layers.
Clearly, the appearance of this structure is due to the azimuthal modulation.
If the value of ${{w}_{0}}$ is small enough, the azimuthal modulation will results in a very narrow structure,
which is quite similar to a one-dimensional finite-energy Airy beam (viz. $w_0\rightarrow0$).
In fact, one can use Dirac's delta fuction
\[
\delta(\theta-\theta_0)=\lim_{w_0\rightarrow 0}\exp \left( -\frac{{{(\theta -{{\theta }_{0}})}^{2}}}{w_{0}^{2}} \right)=
\left\{
\begin{array}{ll}
1,&\theta=\theta_0\\
0,&\theta\neq\theta_0
\end{array}
\right.
\]
to transform the CAi beam in Eq. (\ref{eq5}) into the one-dimensional finite-energy Airy beam.

In Fig. \ref{fig1}(c), we show such a structure by choosing ${{w}_{0}}=0.05$ and $\theta_0=0$.
For the azimuthal modulation along other azimuthal angles,
one can rotate the transverse axes and obtain the same case as for ${{\theta }_{0}}= 0$, by using the  rotation matrix
\[
\left[
\begin{matrix}
   {{x}_{p}}  \\
   {{y}_{p}}  \\
\end{matrix}
\right]
=
\left[
\begin{matrix}
   \cos {{\theta }_{0}} & -\sin {{\theta }_{0}}  \\
   \sin {{\theta }_{0}} & \cos {{\theta }_{0}}  \\
\end{matrix}
\right]
\left[
\begin{matrix}
   x  \\
   y  \\
\end{matrix}
\right],
\]
where $x_p$ and $y_p$ are the coordinates after rotation.
One can verify that the positive $x$ direction after rotation is overlapping with the azimuthal modulation along $\theta_0$, because
\[
\tan {\theta _p} = \frac{{{y_p}}}{{{x_p}}} = \frac{{x\sin {\theta _0} + y\cos {\theta _0}}}{{x\cos {\theta _0} - y\sin {\theta _0}}} = \tan (\theta  + {\theta _0}).
\]
As a result, the CAi beam can be approximately viewed as an assembly of one-dimensional finite-energy Airy beams
associated with the transformed coordinate $x_p$ (which is the variable of the transformed Airy beam), of the form
\begin{equation}\label{eq6}
\psi (x,y) = \sum\limits_{\theta _0 =  - \pi }^{ + \pi } {\rm{Ai}}(r_0 - x_p)\exp [a(r_0 - x_p)].
\end{equation}
In Fig. \ref{fig1}(d), we choose 40 azimuthally modulated CAi beams along different angles, to mimic the initial CAi beam shown in Fig. \ref{fig1}(a).
One can see that such an approximation makes sense.

Therefore, the description of the propagation of a CAi beam can be approximately reduced to the one-dimensional case \cite{zhang.wulixuebao.62.34209.2013},
i.e., to Eq. (\ref{eq3}), which is a much simpler problem.
Thus, the propagation of the CAi beam can be described by
\begin{align}\label{eq7}
\psi (x,y,z) = C & \sum\limits_{\theta _0 =  - \pi }^{ + \pi } {\rm Ai}\left[iaz+\left(\frac{1}{2}f_1-\frac{1}{4}z^2+(r_0-x_p)\right)\right] \times \notag \\
             & \exp\left[a\left(\frac{1}{2}f_1-\frac{1}{2}z^2+(r_0-x_p)\right)\right]\times \notag\\
             & \exp\left[i\left(\frac{1}{2}a^2z+\frac{1}{4}zf_1-\frac{1}{8}f_2-\frac{1}{12}z^3-\frac{1}{2}(r_0-x_p)g+\frac{1}{2}(r_0-x_p)z\right)\right] ,
\end{align}
by analogy to the one-dimensional case \cite{efremidis.ol.36.3006.2011}.
In Eq. (\ref{eq7}),
\[
g(z)=\int_0^z d(t) dt,
\]
\[
f_1(z)=f_0+\int_0^z g(t) dt,
\]
and
\[
f_2(z)=\int_0^z g^2(t) dt,
\]
where $f_0$ is a constant that guarantees $f_1(z=0)=0$.
In addition, parameter $C$ in Eq. (\ref{eq7}) is a normalization parameter,
which makes a balance between the total intensity of components and the intensity of the CAi beam.
According to Eq. (\ref{eq7}), the trajectory of each component in a dynamic linear potential is
\begin{equation}\label{eq8}
  x_p=r_0+\frac{1}{2}f_1-\frac{1}{4}z^2,
\end{equation}
which is related to the type of the potential.
In other words, the CAi beam can be effectively manipulated by the linear potential.
It is clear that the accuracy of our approximation improves as the summation in Eq. (\ref{eq7}) is performed over more terms.
We stress the fact that the solution in Eq. (\ref{eq7}) can be viewed as a superposition of infinitely many one-dimensional Airy beams from different directions,
which is an extension of the work done in \cite{zhang.oe.22.7160.2014}.
In addition, we also have to mention that the summation in Eq. (\ref{eq7}) and the trajectory in Eq. (\ref{eq8}) are somewhat rude approximations,
because we omitted the absolute value sign in Eq. (\ref{eq3}) in the analysis, so that the results for components
displayed in Eqs. (\ref{eq7}) and (\ref{eq8}) will not be very accurate in case of collisions during propagation.
However, such inaccuracy will be waived if the components do not collide with each other.

 If $d(z)\equiv0$, Eq. (\ref{eq1}) reduces to the Schr\"odinger equation in free space
and the propagation of CAi beam according to such a model has already been reported \cite{efremidis.ol.35.4045.2010}.
Equation (\ref{eq7}) can describe the autofocusing effect effectively.
As a test, one can choose only two components with $\theta_0=0$ and $\pi$ to mimic the autofocusing effect,
which is similar to Fig. 2 in \cite{zhang.oe.22.7160.2014}.
If there is no autofocusing during propagation, Eq. (\ref{eq7}) can be rewritten as
\begin{align}\label{eq9}
\psi (x,y,z) = & {\rm Ai}\left[iaz+\left(\frac{1}{2}f_1-\frac{1}{4}z^2+(r_0-r)\right)\right] \times \notag \\
             & \exp\left[a\left(\frac{1}{2}f_1-\frac{1}{2}z^2+(r_0-r)\right)\right]\times \notag\\
             & \exp\left[i\left(\frac{1}{2}a^2z+\frac{1}{4}zf_1-\frac{1}{8}f_2-\frac{1}{12}z^3-\frac{1}{2}(r_0-r)g+\frac{1}{2}(r_0-r)z\right)\right] ,
\end{align}
which is invalid when ``autofocusing'' happens.
For accuracy, the focusing is now guided by the linear potential, and it is not an automatic behavior;
that's why here (and from now on) we place the quotation marks on the word.
But, to stress, autofocusing is adequately described by Eq. (\ref{eq7}).

In summary, our analysis is based on applying the azimuthal modulation to CAi beams;
this novel treatment helps one reduce the two-dimensional problem into a simpler one-dimensional case.
Using Eqs. (\ref{eq7}) and (\ref{eq9}), the propagation of a CAi beam manipulated by a dynamic linear potential can now be simply described.
In the following, we investigate the influence of different potentials on Airy beams during propagation.

\section{Results and discussions}\label{results}

\subsection{The case $d(z)=1$}
\label{case1}

According to Eq. (\ref{eq8}), the trajectory of each component is
\begin{equation}\label{eq10}
  x_p=r_0,
\end{equation}
which is independent on the propagation distance,
i.e., the size of the ring of the CAi beam will not change during propagation.
Actually, Eq. (\ref{eq3}) with $d(z)=1$ has already been solved in \cite{liu.ol.36.1164.2011},
in which the trajectory of the one-dimensional finite-energy Airy beam is a straight line.
Since the beam will not focus during propagation,
Eq. (\ref{eq9}) can be used to describe the propagation.

\begin{figure}[htbp]
\centering
  \includegraphics[width=0.6\columnwidth]{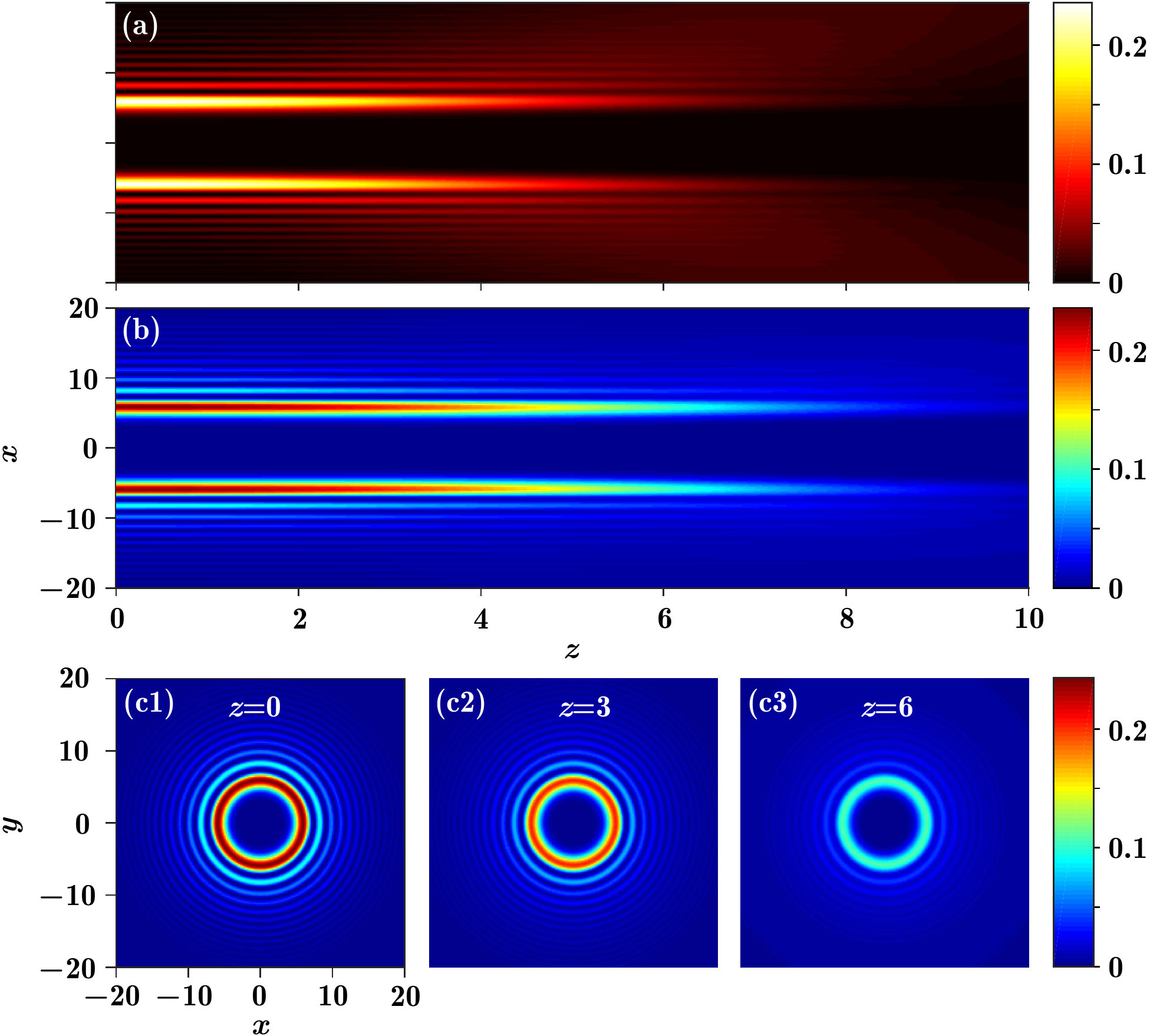}
  \caption{(a) Analytical intensity distribution of a CAi beam during propagation in the $x-z$ plane at $y=0$, according to Eq. (\ref{eq9}), for $d=1$.
  (b) Same as (a), but for numerical simulation.
  (c) Intensity profiles of CAi beam at different propagation distances (noted in each panel).
  Other parameters: ${{r}_{0}}=5$ and $a=0.1$.
  All the panels in (c) share the same variables, dimensions, and color scales.
  }
  \label{fig2}
\end{figure}

The result is displayed in Fig. \ref{fig2},
in which Fig. \ref{fig2}(a) is the analytical result according to Eq. (\ref{eq9}),
and Fig. \ref{fig2}(b) is the corresponding numerical simulation of Eq. (\ref{eq1}).
Due to rotational symmetry of the CAi beam, we only exhibit the
transverse intensity distribution along the $x$ axis ($y=0$) during propagation.
One can see that the analytical and numerical results agree with each other very well.
To elucidate propagation, we also present beam intensities obtained numerically at certain distances,
as displayed in Figs. \ref{fig2}(c1)-\ref{fig2}(c3).
Indeed, the size of the ring of the CAi beam remains the same during propagation, but the intensity dims.
For this case, the linear potential exerts a ``pulling'' influence
that can balance the virtual force which makes the beam focus.

\begin{figure}[htbp]
\centering
  \includegraphics[width=0.6\columnwidth]{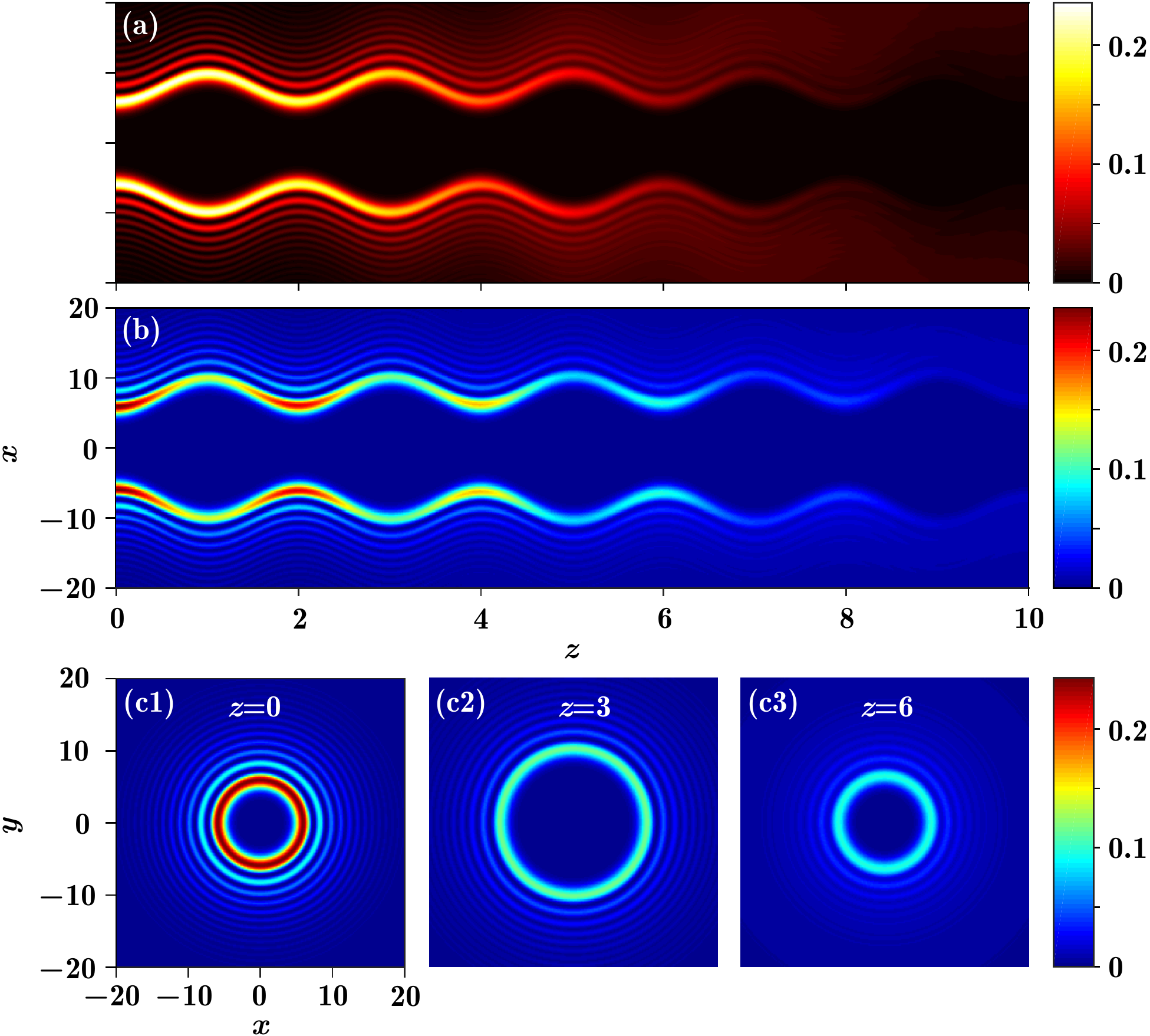}
  \caption{Same as Fig. \ref{fig2}, but for $d(z)=1+4\pi^2\cos(\pi z)$.
  }
  \label{fig3}
\end{figure}

\subsection{The case $d(z)=1+4\pi^2\cos(\pi z)$}\label{case2}

For this case, the trajectory of each component can be written as
\begin{equation}\label{eq11}
  x_p=r_0-2\cos(\pi z)+2,
\end{equation}
which is same as the one displayed in \cite{efremidis.ol.36.3006.2011} with the period of 2,
and the trajectory of the one-dimensional finite-energy Airy beam does not move across $x_p=0$.
So, there will be no ``autofocusing'' of the CAi beam,
which means that the linear potential also exerts a ``pulling'' effect.
As a result, the propagation for this case can still be described by Eq. (\ref{eq9}).
The corresponding result is exhibited in Fig. \ref{fig3},
the setup of which is the same as that of Fig. \ref{fig2}.
Again, the analytical and numerical results agree with each other very well.

\subsection{The case $d(z)=13-12z$}\label{case3}

\begin{figure}[htbp]
\centering
  \includegraphics[width=0.6\columnwidth]{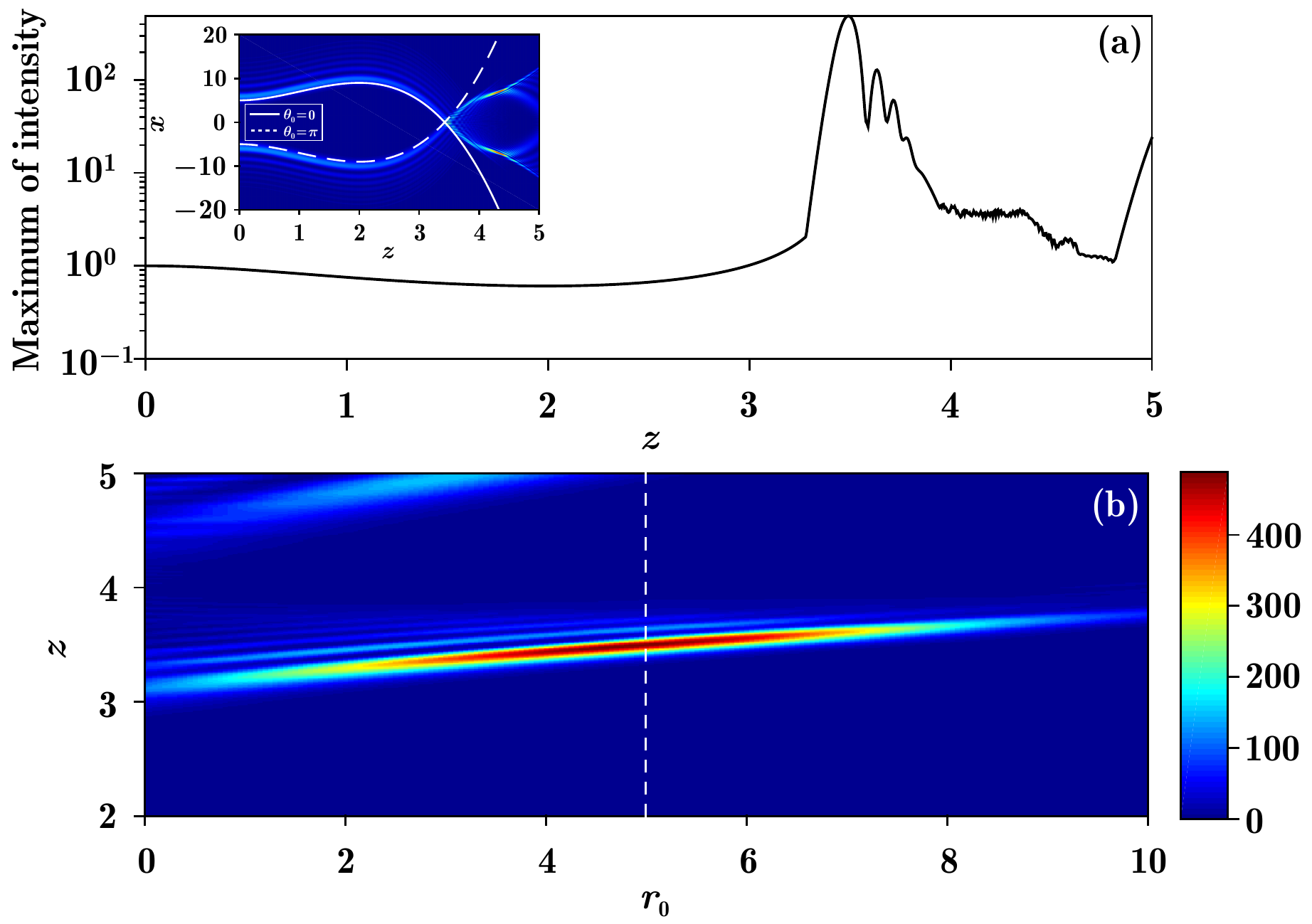}
  \caption{(a) Maximum of the beam intensity during propagation for $d(z)=13-12z$ and $r_0=5$.
  The maximum of the beam intensity at $z=0$ is 1.
  Inset shows the propagation of the components corresponding to $\theta_0=0$ and $\theta_0=\pi$, and the corresponding theoretical trajectories.
  (b) The maximum of the beam intensity as a function of $r_0$ and $z$.
  The white dashed line corresponds to the curve in (a).
  The decay parameter is $a=0.05$.
  }
  \label{fig4}
\end{figure}

According to Eq. (\ref{eq8}), the trajectory for this case is
\begin{equation}\label{eq12}
x_p = r_0 - z^3 + 3 z^2,
\end{equation}
which indicates that the CAi beam will undergo ``autofocusing'' during propagation,
so that the description of the propagation should be based on Eq. (\ref{eq7}).
In Fig. \ref{fig4}(a), the maximum of intensity during propagation is exhibited.
In order to make a comparison, we take the same values for $a$ as used in \cite{efremidis.ol.35.4045.2010}
and the same normalized intensity of the input.
One can clearly see that the beam intensity decreases initially, but then increases to almost 500 after ``autofocusing'',
which is much higher than that without a potential.
Undoubtedly, the potential exerts a strong ``pushing'' effect that strengthens the focusing.

To show the ``pushing'' effect more clearly, we display the propagation
of the components corresponding to $\theta_0=0$ and $\theta_0=\pi$ in the inset in Fig. \ref{fig4}(a).
It is evident that at first the two components tend to separate before the focusing point, but then they converge.
The slope of the components at the colliding point is bigger than that without a potential, as in \cite{efremidis.ol.35.4045.2010},
which can be viewed that the components acquire a larger speed because of the ``pushing'' effect; hence, the ``autofocusing'' is strengthened.
However, according to the propagation in the inset, one may doubt the accuracy of our analysis, because the intensity at the ``autofocusing'' point is not the largest.
The explanation is that one cannot use only two one-dimensional Airy beams to obtain the real physical picture of the propagation of the full CAi beam
-- such an approximation is too poor.
If another pair of components is added, the intensity at and only at the ``autofocusing'' point will be larger, due to the collision of four components.
The intensity at the ``autofocusing'' point will reach an enormous value when numerous pairs of components are considered simultaneously;
such a consideration can approximate the propagation of the CAi beam very well.
Since the theoretical trajectories in Eq. (\ref{eq8}) do not consider the absolute value sign in Eq. (\ref{eq3}),
the theoretical trajectories are only valid before collision occurs.

Similar to the previous study \cite{efremidis.ol.35.4045.2010}, $r_0$ also affects the maximum of the beam intensity (MBI) reached during propagation.
To observe such a dependence more clearly, we show the change of the MBI as a function of both $r_0$ and the propagation distance, as displayed in Fig. \ref{fig4}(b).
One sees that the MBI first increases and then decreases with the increasing of $r_0$,
and the MBI reaches a maximum around $r_0=5$.
In addition, one can also see that the location of MBI also changes with $r_0$.
The reason is that the ``autofocusing'' effect requires a longer distance to establish itself when $r_0$ increases.

\subsection{The case $d(z)=1-4H(z-2)/\left[ (z-4)z+5 \right]^{3/2}$}\label{case4}

\begin{figure}[htbp]
\centering
  \includegraphics[width=0.6\columnwidth]{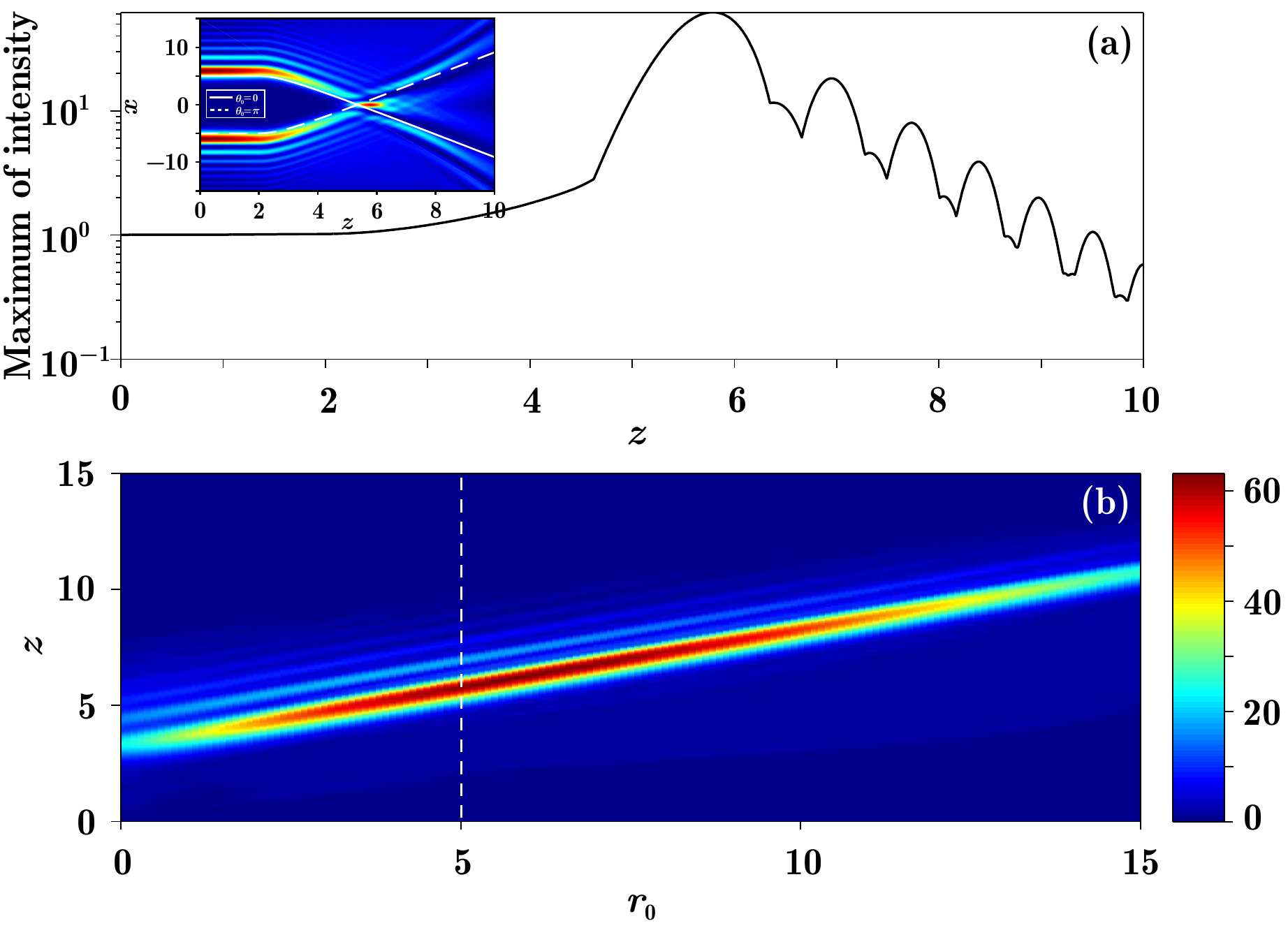}
  \caption{Figure setup is as in Fig. \ref{fig4}.
  }
  \label{fig5}
\end{figure}

The function $H(z-2)$ here is the Heaviside step function, which demands
\begin{equation*}
H(z-2)=
\left\{
\begin{array}{ll}
0,   &{\rm for}~(z\leq2),\\
1,   &{\rm for}~(z>2).
\end{array}
\right.
\end{equation*}
Therefore, the trajectory for this case can be written as
\begin{equation}\label{eq13}
x_p=
\left\{
\begin{array}{ll}
r_0,   ~&{\rm for}~(z\leq2),\\
r_0+2-2\sqrt{(z-4)z+5},   ~&{\rm for}~(z>2).
\end{array}
\right.
\end{equation}
Similar to the case in Sec. \ref{case3}, the beam will ``autofocus'' during propagation,
and we display the MBI for this case in Fig. \ref{fig5}.
In Fig. \ref{fig5}(a), one finds that the MBI during propagation is about 60 times of the input.
This value is much lower than that in \cite{efremidis.ol.35.4045.2010}, as well as of that in Sec. \ref{case3}.
From the inset in Fig. \ref{fig5}(a), one finds the reason easily --
the speed (slope) at the colliding point is smaller than that in \cite{efremidis.ol.35.4045.2010}, because of the ``pulling'' effect.
Figure \ref{fig5}(b) shows the MBI as a function of $r_0$ and $z$, from which one can see that the peak of the MBI
is still no more than 65, so the linear potential in this case exerts a ``pulling'' influence,
which weakens the autofocusing effect.

\section{Circular Airy beam with Gaussian apodization}\label{gauss}

By using a Gaussian apodization, the CAi beam can also be  made finite-energy, of the form
\[
\psi={\rm Ai}(r_0-r)\exp\left(-\frac{(r_0-r)^2}{w^2}\right),
\]
where $w$ is related to the beam width of the Gaussian apodization.
Following the same procedure as before, the solution can be written as
\begin{align}\label{eq13}
\psi (x,y,z)= & \frac{1}{\sqrt \sigma} \sum\limits_{\theta_0=-\pi }^{+\pi} {\rm Ai} \left[ \frac{1}{\sigma} \left((r_0-x_p)+\frac{f_1}{2}\right) - \frac{z^2}{4\sigma^2} \right]
 \exp \left[ \frac{-1}{\sigma w^2} \left( (r_0-x_p)+\frac{f_1}{2}\right) \right]\times  \notag\\
 & \exp \left\{ i\left[ -\frac{f_2}{8}-\frac{D}{2}(r_0-x_p)+\frac{z}{2\sigma^2 }\left( r_0-x_p+\frac{f_1}{2}-\frac{z^2}{6\sigma} \right) \right] \right\},
\end{align}	
where $\sigma =1+{2iz}/{w^2}$.
If there is no ``autofocusing'' during propagation, the solution in Eq. (\ref{eq13}) can be directly rewritten as
\begin{align}\label{eq14}
\psi (x,y,z)= & \frac{1}{\sqrt \sigma} {\rm Ai} \left[ \frac{1}{\sigma} \left((r_0-r)+\frac{f_1}{2}\right) - \frac{z^2}{4\sigma^2} \right]
 \exp \left[ \frac{-1}{\sigma w^2} \left( (r_0-r)+\frac{f_1}{2}\right) \right]\times  \notag\\
 & \exp \left\{ i\left[ -\frac{f_2}{8}-\frac{D}{2}(r_0-r)+\frac{z}{2\sigma^2 }\left( r_0-r+\frac{f_1}{2}-\frac{z^2}{6\sigma} \right) \right] \right\}.
\end{align}	
Clearly, the trajectory of each component can be written as
\begin{equation}\label{eq15}
{{x}_{p}}={{r}_{0}}-\frac{{{z}^{2}}{{w}^{4}}}{4({{w}^{4}}+4{{z}^{2}})}+\frac{{{f}_{1}}}{2},
\end{equation}
which is not only related to the linear potential, but also to the beam width of the Gaussian apodization.
If the condition $w^2\gg z$ is fulfilled (this condition can be relatively easily fulfilled when the propagation distance is not too long),
Eq. (\ref{eq15}) will be reduced to Eq. (\ref{eq8});
therefore, all the phenomena obtained in Sec. \ref{results} can also be realized for Gaussian apodization.
In other words, the Gaussian-apodized CAi beams can also be well controlled and manipulated.
However, one needs to bear in mind that such an apodization introduces another degree of freedom in the beam manipulation.

\begin{figure}[htbp]
\centering
  \includegraphics[width=0.6\columnwidth]{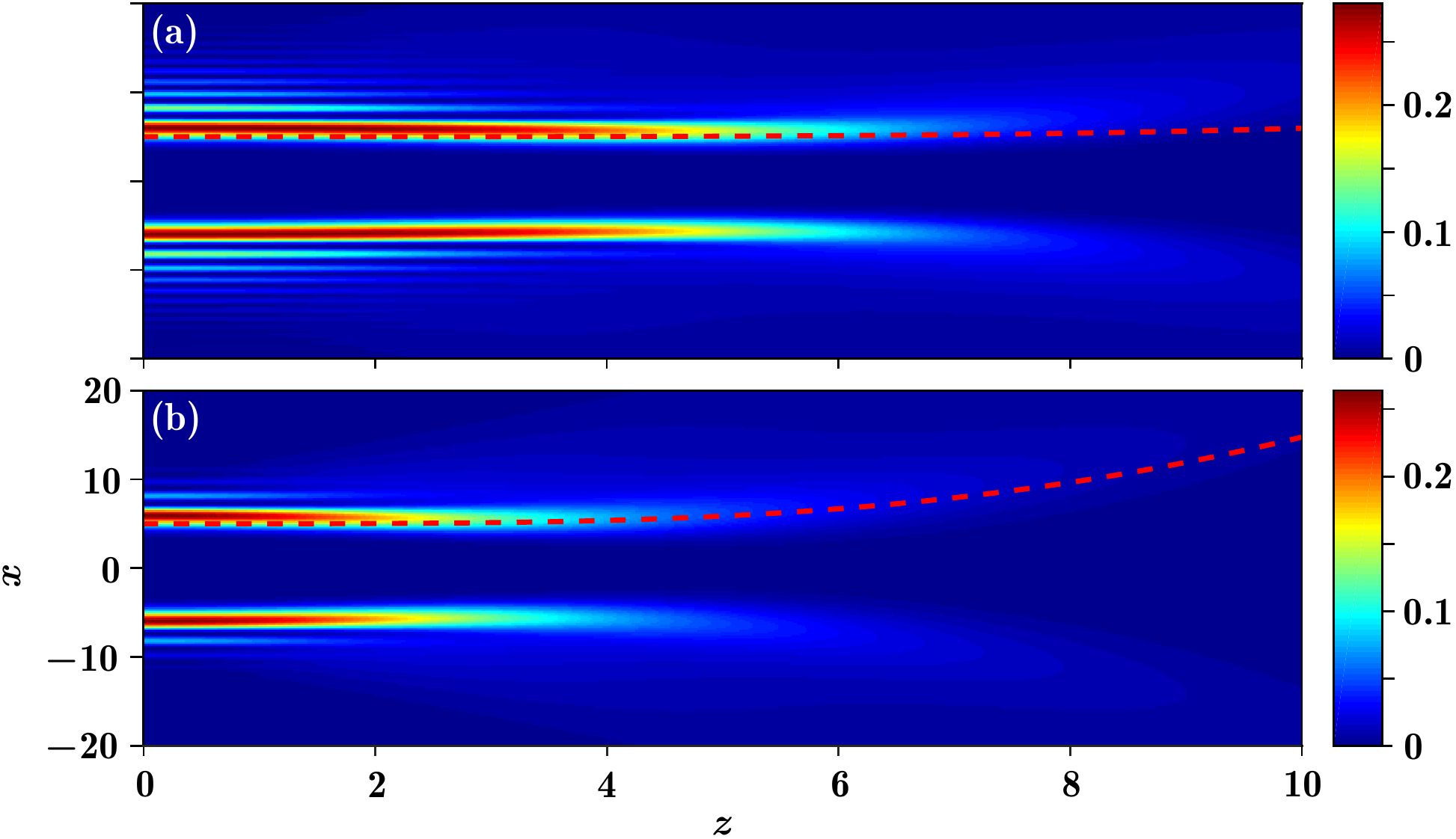}
  \caption{Same as Fig. \ref{fig2}(b), but for Gaussian apodization with
  (a) $w=10$, and (b) $w=5$.
  The dashed curves represent the analytical trajectories.
  The panels share the same variables and dimensions.
  }
  \label{fig6}
\end{figure}

As an elucidation, in Fig. \ref{fig6}
we display the numerical simulations of a CAi beam with Gaussian apodization in a dynamic linear potential that is used in Sec. \ref{case1};
the dashed curves show the corresponding analytical trajectories.
One can find that, in comparison with the trajectory in Fig. \ref{fig6}(b), the trajectory in Fig. \ref{fig6}(a) is much more similar to that in Fig. \ref{fig2}(b),
because the value of $w$ in the latter case is bigger, which guarantees the validity of the condition $w^2\gg z$ over a longer propagation distance.

%

\section{Conclusion}
\label{conclusion}

In summary, we have theoretically and numerically investigated the spatial guidance of CAi beams, by using different types of dynamic linear potentials.
We find that the accelerating trajectory of the CAi beam can be controlled by the linear potential effectively.
Such a manipulation may weaken (even eliminate) or strengthen the autofocusing effect of the CAi beam, depending on the form of the linear potential.
Our results broaden the potential applications of CAi beams in trapping and manipulating microparticles, offering potential use
in optics, biology, and other disciplines.

Last but not least, we would like to point out that the analysis presented here can also be applied
to the outward CAi beams \cite{hwang.oe.19.7356.2011}, but the validity is more limited,
because the outward CAi beams always produce a focusing peak at the beam's center,
which is not the case for the inward CAi beams if the linear potential is well chosen.
As reported in \cite{hwang.pj.4.174.2012}, the manipulation on the ``autofocusing'' effect can be enriched if one launches the input beam appropriately (e.g., the incident angle),
which is independent of the potential.

\section*{Acknowledgment}
This work was supported by the 973 Program (2012CB921804),
KSTIT of Shaanxi province (2014KCT-10),
NSFC (11474228, 61308015),
and the NPRP 6-021-1-005 project of the Qatar National Research Fund (a member of the Qatar Foundation).
MRB also acknowledges support by the Al Sraiya Holding Group.

\bibliographystyle{myprx}
\bibliography{my_refs_library}

\begin{thebibliography}{42}%
\makeatletter
\providecommand \@ifxundefined [1]{%
 \@ifx{#1\undefined}
}%
\providecommand \@ifnum [1]{%
 \ifnum #1\expandafter \@firstoftwo
 \else \expandafter \@secondoftwo
 \fi
}%
\providecommand \@ifx [1]{%
 \ifx #1\expandafter \@firstoftwo
 \else \expandafter \@secondoftwo
 \fi
}%
\providecommand \natexlab [1]{#1}%
\providecommand \enquote  [1]{``#1''}%
\providecommand \bibnamefont  [1]{#1}%
\providecommand \bibfnamefont [1]{#1}%
\providecommand \citenamefont [1]{#1}%
\providecommand \href@noop [0]{\@secondoftwo}%
\providecommand \href [0]{\begingroup \@sanitize@url \@href}%
\providecommand \@href[1]{\@@startlink{#1}\@@href}%
\providecommand \@@href[1]{\endgroup#1\@@endlink}%
\providecommand \@sanitize@url [0]{\catcode `\\12\catcode `\$12\catcode
  `\&12\catcode `\#12\catcode `\^12\catcode `\_12\catcode `\%12\relax}%
\providecommand \@@startlink[1]{}%
\providecommand \@@endlink[0]{}%
\providecommand \url  [0]{\begingroup\@sanitize@url \@url }%
\providecommand \@url [1]{\endgroup\@href {#1}{\urlprefix }}%
\providecommand \urlprefix  [0]{URL }%
\providecommand \Eprint [0]{\href }%
\providecommand \doibase [0]{http://dx.doi.org/}%
\providecommand \selectlanguage [0]{\@gobble}%
\providecommand \bibinfo  [0]{\@secondoftwo}%
\providecommand \bibfield  [0]{\@secondoftwo}%
\providecommand \translation [1]{[#1]}%
\providecommand \BibitemOpen [0]{}%
\providecommand \bibitemStop [0]{}%
\providecommand \bibitemNoStop [0]{.\EOS\space}%
\providecommand \EOS [0]{\spacefactor3000\relax}%
\providecommand \BibitemShut  [1]{\csname bibitem#1\endcsname}%
\let\auto@bib@innerbib\@empty
\bibitem [{\citenamefont {Berry}\ and\ \citenamefont
  {Balazs}(1979)}]{berry.ajp.47.264.1979}%
  \BibitemOpen
  \bibfield  {author} {\bibinfo {author} {\bibfnamefont {M.~V.}\ \bibnamefont
  {Berry}}\ and\ \bibinfo {author} {\bibfnamefont {N.~L.}\ \bibnamefont
  {Balazs}},\ }\emph {\bibinfo {title} {Nonspreading wave packets}},\ \href
  {\doibase 10.1119/1.11855} {\bibfield  {journal} {\bibinfo  {journal} {Am. J.
  Phys.}\ }\textbf {\bibinfo {volume} {47}},\ \bibinfo {pages} {264} (\bibinfo
  {year} {1979})}\BibitemShut {NoStop}%
\bibitem [{\citenamefont {Siviloglou}\ and\ \citenamefont
  {Christodoulides}(2007)}]{siviloglou.ol.32.979.2007}%
  \BibitemOpen
  \bibfield  {author} {\bibinfo {author} {\bibfnamefont {G.~A.}\ \bibnamefont
  {Siviloglou}}\ and\ \bibinfo {author} {\bibfnamefont {D.~N.}\ \bibnamefont
  {Christodoulides}},\ }\emph {\bibinfo {title} {Accelerating finite energy
  {A}iry beams}},\ \href {\doibase 10.1364/OL.32.000979} {\bibfield  {journal}
  {\bibinfo  {journal} {Opt. Lett.}\ }\textbf {\bibinfo {volume} {32}},\
  \bibinfo {pages} {979} (\bibinfo {year} {2007})}\BibitemShut {NoStop}%
\bibitem [{\citenamefont {Siviloglou}\ \emph {et~al.}(2007)\citenamefont
  {Siviloglou}, \citenamefont {Broky}, \citenamefont {Dogariu},\ and\
  \citenamefont {Christodoulides}}]{siviloglou.prl.99.213901.2007}%
  \BibitemOpen
  \bibfield  {author} {\bibinfo {author} {\bibfnamefont {G.}~\bibnamefont
  {Siviloglou}}, \bibinfo {author} {\bibfnamefont {J.}~\bibnamefont {Broky}},
  \bibinfo {author} {\bibfnamefont {A.}~\bibnamefont {Dogariu}}, \ and\
  \bibinfo {author} {\bibfnamefont {D.}~\bibnamefont {Christodoulides}},\
  }\emph {\bibinfo {title} {Observation of Accelerating {A}iry Beams}},\ \href
  {\doibase 10.1103/PhysRevLett.99.213901} {\bibfield  {journal} {\bibinfo
  {journal} {Phys. Rev. Lett.}\ }\textbf {\bibinfo {volume} {99}},\ \bibinfo
  {pages} {213901} (\bibinfo {year} {2007})}\BibitemShut {NoStop}%
\bibitem [{\citenamefont {Broky}\ \emph {et~al.}(2008)\citenamefont {Broky},
  \citenamefont {Siviloglou}, \citenamefont {Dogariu},\ and\ \citenamefont
  {Christodoulides}}]{broky.oe.16.12880.2008}%
  \BibitemOpen
  \bibfield  {author} {\bibinfo {author} {\bibfnamefont {J.}~\bibnamefont
  {Broky}}, \bibinfo {author} {\bibfnamefont {G.~A.}\ \bibnamefont
  {Siviloglou}}, \bibinfo {author} {\bibfnamefont {A.}~\bibnamefont {Dogariu}},
  \ and\ \bibinfo {author} {\bibfnamefont {D.~N.}\ \bibnamefont
  {Christodoulides}},\ }\emph {\bibinfo {title} {Self-healing properties of
  optical {A}iry beams}},\ \href {\doibase 10.1364/OE.16.012880} {\bibfield
  {journal} {\bibinfo  {journal} {Opt. Express}\ }\textbf {\bibinfo {volume}
  {16}},\ \bibinfo {pages} {12880} (\bibinfo {year} {2008})}\BibitemShut
  {NoStop}%
\bibitem [{\citenamefont {Kaminer}\ \emph {et~al.}(2011)\citenamefont
  {Kaminer}, \citenamefont {Segev},\ and\ \citenamefont
  {Christodoulides}}]{kaminer.prl.106.213903.2011}%
  \BibitemOpen
  \bibfield  {author} {\bibinfo {author} {\bibfnamefont {I.}~\bibnamefont
  {Kaminer}}, \bibinfo {author} {\bibfnamefont {M.}~\bibnamefont {Segev}}, \
  and\ \bibinfo {author} {\bibfnamefont {D.~N.}\ \bibnamefont
  {Christodoulides}},\ }\emph {\bibinfo {title} {Self-Accelerating Self-Trapped
  Optical Beams}},\ \href {\doibase 10.1103/PhysRevLett.106.213903} {\bibfield
  {journal} {\bibinfo  {journal} {Phys. Rev. Lett.}\ }\textbf {\bibinfo
  {volume} {106}},\ \bibinfo {pages} {213903} (\bibinfo {year}
  {2011})}\BibitemShut {NoStop}%
\bibitem [{\citenamefont {Lotti}\ \emph {et~al.}(2011)\citenamefont {Lotti},
  \citenamefont {Faccio}, \citenamefont {Couairon}, \citenamefont {Papazoglou},
  \citenamefont {Panagiotopoulos}, \citenamefont {Abdollahpour},\ and\
  \citenamefont {Tzortzakis}}]{lotti.pra.84.021807.2011}%
  \BibitemOpen
  \bibfield  {author} {\bibinfo {author} {\bibfnamefont {A.}~\bibnamefont
  {Lotti}}, \bibinfo {author} {\bibfnamefont {D.}~\bibnamefont {Faccio}},
  \bibinfo {author} {\bibfnamefont {A.}~\bibnamefont {Couairon}}, \bibinfo
  {author} {\bibfnamefont {D.~G.}\ \bibnamefont {Papazoglou}}, \bibinfo
  {author} {\bibfnamefont {P.}~\bibnamefont {Panagiotopoulos}}, \bibinfo
  {author} {\bibfnamefont {D.}~\bibnamefont {Abdollahpour}}, \ and\ \bibinfo
  {author} {\bibfnamefont {S.}~\bibnamefont {Tzortzakis}},\ }\emph {\bibinfo
  {title} {Stationary nonlinear {A}iry beams}},\ \href {\doibase
  10.1103/PhysRevA.84.021807} {\bibfield  {journal} {\bibinfo  {journal} {Phys.
  Rev. A}\ }\textbf {\bibinfo {volume} {84}},\ \bibinfo {pages} {021807}
  (\bibinfo {year} {2011})}\BibitemShut {NoStop}%
\bibitem [{\citenamefont {Dolev}\ \emph {et~al.}(2012)\citenamefont {Dolev},
  \citenamefont {Kaminer}, \citenamefont {Shapira}, \citenamefont {Segev},\
  and\ \citenamefont {Arie}}]{dolev.prl.108.113903.2012}%
  \BibitemOpen
  \bibfield  {author} {\bibinfo {author} {\bibfnamefont {I.}~\bibnamefont
  {Dolev}}, \bibinfo {author} {\bibfnamefont {I.}~\bibnamefont {Kaminer}},
  \bibinfo {author} {\bibfnamefont {A.}~\bibnamefont {Shapira}}, \bibinfo
  {author} {\bibfnamefont {M.}~\bibnamefont {Segev}}, \ and\ \bibinfo {author}
  {\bibfnamefont {A.}~\bibnamefont {Arie}},\ }\emph {\bibinfo {title}
  {Experimental Observation of Self-Accelerating Beams in Quadratic Nonlinear
  Media}},\ \href {\doibase 10.1103/PhysRevLett.108.113903} {\bibfield
  {journal} {\bibinfo  {journal} {Phys. Rev. Lett.}\ }\textbf {\bibinfo
  {volume} {108}},\ \bibinfo {pages} {113903} (\bibinfo {year}
  {2012})}\BibitemShut {NoStop}%
\bibitem [{\citenamefont {Zhang}\ \emph
  {et~al.}(2013{\natexlab{a}})\citenamefont {Zhang}, \citenamefont {Beli\'{c}},
  \citenamefont {Wu}, \citenamefont {Zheng}, \citenamefont {Lu}, \citenamefont
  {Li},\ and\ \citenamefont {Zhang}}]{zhang.ol.38.4585.2013}%
  \BibitemOpen
  \bibfield  {author} {\bibinfo {author} {\bibfnamefont {Y.~Q.}\ \bibnamefont
  {Zhang}}, \bibinfo {author} {\bibfnamefont {M.}~\bibnamefont {Beli\'{c}}},
  \bibinfo {author} {\bibfnamefont {Z.~K.}\ \bibnamefont {Wu}}, \bibinfo
  {author} {\bibfnamefont {H.~B.}\ \bibnamefont {Zheng}}, \bibinfo {author}
  {\bibfnamefont {K.~Q.}\ \bibnamefont {Lu}}, \bibinfo {author} {\bibfnamefont
  {Y.~Y.}\ \bibnamefont {Li}}, \ and\ \bibinfo {author} {\bibfnamefont {Y.~P.}\
  \bibnamefont {Zhang}},\ }\emph {\bibinfo {title} {Soliton pair generation in
  the interactions of {A}iry and nonlinear accelerating beams}},\ \href
  {\doibase 10.1364/OL.38.004585} {\bibfield  {journal} {\bibinfo  {journal}
  {Opt. Lett.}\ }\textbf {\bibinfo {volume} {38}},\ \bibinfo {pages} {4585}
  (\bibinfo {year} {2013}{\natexlab{a}})}\BibitemShut {NoStop}%
\bibitem [{\citenamefont {Zhang}\ \emph {et~al.}(2014)\citenamefont {Zhang},
  \citenamefont {Beli\'{c}}, \citenamefont {Zheng}, \citenamefont {Chen},
  \citenamefont {Li}, \citenamefont {Li},\ and\ \citenamefont
  {Zhang}}]{zhang.oe.22.7160.2014}%
  \BibitemOpen
  \bibfield  {author} {\bibinfo {author} {\bibfnamefont {Y.~Q.}\ \bibnamefont
  {Zhang}}, \bibinfo {author} {\bibfnamefont {M.~R.}\ \bibnamefont
  {Beli\'{c}}}, \bibinfo {author} {\bibfnamefont {H.~B.}\ \bibnamefont
  {Zheng}}, \bibinfo {author} {\bibfnamefont {H.~X.}\ \bibnamefont {Chen}},
  \bibinfo {author} {\bibfnamefont {C.~B.}\ \bibnamefont {Li}}, \bibinfo
  {author} {\bibfnamefont {Y.~Y.}\ \bibnamefont {Li}}, \ and\ \bibinfo {author}
  {\bibfnamefont {Y.~P.}\ \bibnamefont {Zhang}},\ }\emph {\bibinfo {title}
  {Interactions of {A}iry beams, nonlinear accelerating beams, and induced
  solitons in {K}err and saturable nonlinear media}},\ \href {\doibase
  10.1364/OE.22.007160} {\bibfield  {journal} {\bibinfo  {journal} {Opt.
  Express}\ }\textbf {\bibinfo {volume} {22}},\ \bibinfo {pages} {7160}
  (\bibinfo {year} {2014})}\BibitemShut {NoStop}%
\bibitem [{\citenamefont {Shen}\ \emph {et~al.}(2015)\citenamefont {Shen},
  \citenamefont {Gao},\ and\ \citenamefont {Ge}}]{shen.sr.5.9814.2015}%
  \BibitemOpen
  \bibfield  {author} {\bibinfo {author} {\bibfnamefont {M.}~\bibnamefont
  {Shen}}, \bibinfo {author} {\bibfnamefont {J.}~\bibnamefont {Gao}}, \ and\
  \bibinfo {author} {\bibfnamefont {L.}~\bibnamefont {Ge}},\ }\emph {\bibinfo
  {title} {Solitons shedding from Airy beams and bound states of breathing
  {A}iry solitons in nonlocal nonlinear media}},\ \href {\doibase
  10.1038/srep09814} {\bibfield  {journal} {\bibinfo  {journal} {Sci. Rep.}\
  }\textbf {\bibinfo {volume} {5}},\ \bibinfo {pages} {9814} (\bibinfo {year}
  {2015})}\BibitemShut {NoStop}%
\bibitem [{\citenamefont {Diebel}\ \emph {et~al.}(2015)\citenamefont {Diebel},
  \citenamefont {Boki\'{c}}, \citenamefont {Timotijevi\'{c}}, \citenamefont
  {Savi\'{c}},\ and\ \citenamefont {Denz}}]{diebel.oe.23.24351.2015}%
  \BibitemOpen
  \bibfield  {author} {\bibinfo {author} {\bibfnamefont {F.}~\bibnamefont
  {Diebel}}, \bibinfo {author} {\bibfnamefont {B.~M.}\ \bibnamefont
  {Boki\'{c}}}, \bibinfo {author} {\bibfnamefont {D.~V.}\ \bibnamefont
  {Timotijevi\'{c}}}, \bibinfo {author} {\bibfnamefont {D.~M.~J.}\ \bibnamefont
  {Savi\'{c}}}, \ and\ \bibinfo {author} {\bibfnamefont {C.}~\bibnamefont
  {Denz}},\ }\emph {\bibinfo {title} {Soliton formation by decelerating
  interacting {A}iry beams}},\ \href {\doibase 10.1364/OE.23.024351} {\bibfield
   {journal} {\bibinfo  {journal} {Opt. Express}\ }\textbf {\bibinfo {volume}
  {23}},\ \bibinfo {pages} {24351} (\bibinfo {year} {2015})}\BibitemShut
  {NoStop}%
\bibitem [{\citenamefont {Efremidis}\ \emph {et~al.}(2013)\citenamefont
  {Efremidis}, \citenamefont {Paltoglou},\ and\ \citenamefont {von
  Klitzing}}]{efremidis.pra.87.043637.2013}%
  \BibitemOpen
  \bibfield  {author} {\bibinfo {author} {\bibfnamefont {N.~K.}\ \bibnamefont
  {Efremidis}}, \bibinfo {author} {\bibfnamefont {V.}~\bibnamefont
  {Paltoglou}}, \ and\ \bibinfo {author} {\bibfnamefont {W.}~\bibnamefont {von
  Klitzing}},\ }\emph {\bibinfo {title} {Accelerating and abruptly autofocusing
  matter waves}},\ \href {\doibase 10.1103/PhysRevA.87.043637} {\bibfield
  {journal} {\bibinfo  {journal} {Phys. Rev. A}\ }\textbf {\bibinfo {volume}
  {87}},\ \bibinfo {pages} {043637} (\bibinfo {year} {2013})}\BibitemShut
  {NoStop}%
\bibitem [{\citenamefont {Salandrino}\ and\ \citenamefont
  {Christodoulides}(2010)}]{salandrino.ol.35.2082.2010}%
  \BibitemOpen
  \bibfield  {author} {\bibinfo {author} {\bibfnamefont {A.}~\bibnamefont
  {Salandrino}}\ and\ \bibinfo {author} {\bibfnamefont {D.~N.}\ \bibnamefont
  {Christodoulides}},\ }\emph {\bibinfo {title} {Airy plasmon: a nondiffracting
  surface wave}},\ \href {\doibase 10.1364/OL.35.002082} {\bibfield  {journal}
  {\bibinfo  {journal} {Opt. Lett.}\ }\textbf {\bibinfo {volume} {35}},\
  \bibinfo {pages} {2082} (\bibinfo {year} {2010})}\BibitemShut {NoStop}%
\bibitem [{\citenamefont {Zhang}\ \emph
  {et~al.}(2011{\natexlab{a}})\citenamefont {Zhang}, \citenamefont {Wang},
  \citenamefont {Liu}, \citenamefont {Yin}, \citenamefont {Lu}, \citenamefont
  {Chen},\ and\ \citenamefont {Zhang}}]{zhang.ol.36.3191.2011}%
  \BibitemOpen
  \bibfield  {author} {\bibinfo {author} {\bibfnamefont {P.}~\bibnamefont
  {Zhang}}, \bibinfo {author} {\bibfnamefont {S.}~\bibnamefont {Wang}},
  \bibinfo {author} {\bibfnamefont {Y.}~\bibnamefont {Liu}}, \bibinfo {author}
  {\bibfnamefont {X.}~\bibnamefont {Yin}}, \bibinfo {author} {\bibfnamefont
  {C.}~\bibnamefont {Lu}}, \bibinfo {author} {\bibfnamefont {Z.}~\bibnamefont
  {Chen}}, \ and\ \bibinfo {author} {\bibfnamefont {X.}~\bibnamefont {Zhang}},\
  }\emph {\bibinfo {title} {Plasmonic {A}iry beams with dynamically controlled
  trajectories}},\ \href {\doibase 10.1364/OL.36.003191} {\bibfield  {journal}
  {\bibinfo  {journal} {Opt. Lett.}\ }\textbf {\bibinfo {volume} {36}},\
  \bibinfo {pages} {3191} (\bibinfo {year} {2011}{\natexlab{a}})}\BibitemShut
  {NoStop}%
\bibitem [{\citenamefont {Minovich}\ \emph {et~al.}(2011)\citenamefont
  {Minovich}, \citenamefont {Klein}, \citenamefont {Janunts}, \citenamefont
  {Pertsch}, \citenamefont {Neshev},\ and\ \citenamefont
  {Kivshar}}]{minovich.prl.107.116802.2011}%
  \BibitemOpen
  \bibfield  {author} {\bibinfo {author} {\bibfnamefont {A.}~\bibnamefont
  {Minovich}}, \bibinfo {author} {\bibfnamefont {A.~E.}\ \bibnamefont {Klein}},
  \bibinfo {author} {\bibfnamefont {N.}~\bibnamefont {Janunts}}, \bibinfo
  {author} {\bibfnamefont {T.}~\bibnamefont {Pertsch}}, \bibinfo {author}
  {\bibfnamefont {D.~N.}\ \bibnamefont {Neshev}}, \ and\ \bibinfo {author}
  {\bibfnamefont {Y.~S.}\ \bibnamefont {Kivshar}},\ }\emph {\bibinfo {title}
  {Generation and Near-Field Imaging of {A}iry Surface Plasmons}},\ \href
  {\doibase 10.1103/PhysRevLett.107.116802} {\bibfield  {journal} {\bibinfo
  {journal} {Phys. Rev. Lett.}\ }\textbf {\bibinfo {volume} {107}},\ \bibinfo
  {pages} {116802} (\bibinfo {year} {2011})}\BibitemShut {NoStop}%
\bibitem [{\citenamefont {Li}\ \emph {et~al.}(2011)\citenamefont {Li},
  \citenamefont {Li}, \citenamefont {Wang}, \citenamefont {Zhang},\ and\
  \citenamefont {Zhu}}]{li.prl.107.126804.2011}%
  \BibitemOpen
  \bibfield  {author} {\bibinfo {author} {\bibfnamefont {L.}~\bibnamefont
  {Li}}, \bibinfo {author} {\bibfnamefont {T.}~\bibnamefont {Li}}, \bibinfo
  {author} {\bibfnamefont {S.~M.}\ \bibnamefont {Wang}}, \bibinfo {author}
  {\bibfnamefont {C.}~\bibnamefont {Zhang}}, \ and\ \bibinfo {author}
  {\bibfnamefont {S.~N.}\ \bibnamefont {Zhu}},\ }\emph {\bibinfo {title}
  {Plasmonic {A}iry Beam Generated by In-Plane Diffraction}},\ \href {\doibase
  10.1103/PhysRevLett.107.126804} {\bibfield  {journal} {\bibinfo  {journal}
  {Phys. Rev. Lett.}\ }\textbf {\bibinfo {volume} {107}},\ \bibinfo {pages}
  {126804} (\bibinfo {year} {2011})}\BibitemShut {NoStop}%
\bibitem [{\citenamefont {Hu}\ \emph {et~al.}(2013)\citenamefont {Hu},
  \citenamefont {Li}, \citenamefont {Bongiovanni}, \citenamefont {Clerici},
  \citenamefont {Yao}, \citenamefont {Chen}, \citenamefont {{Aza\~{n}a}},\ and\
  \citenamefont {Morandotti}}]{hu.ol.38.380.2013}%
  \BibitemOpen
  \bibfield  {author} {\bibinfo {author} {\bibfnamefont {Y.}~\bibnamefont
  {Hu}}, \bibinfo {author} {\bibfnamefont {M.}~\bibnamefont {Li}}, \bibinfo
  {author} {\bibfnamefont {D.}~\bibnamefont {Bongiovanni}}, \bibinfo {author}
  {\bibfnamefont {M.}~\bibnamefont {Clerici}}, \bibinfo {author} {\bibfnamefont
  {J.}~\bibnamefont {Yao}}, \bibinfo {author} {\bibfnamefont {Z.}~\bibnamefont
  {Chen}}, \bibinfo {author} {\bibfnamefont {J.}~\bibnamefont {{Aza\~{n}a}}}, \
  and\ \bibinfo {author} {\bibfnamefont {R.}~\bibnamefont {Morandotti}},\
  }\emph {\bibinfo {title} {Spectrum to distance mapping via nonlinear {A}iry
  pulses}},\ \href {\doibase 10.1364/OL.38.000380} {\bibfield  {journal}
  {\bibinfo  {journal} {Opt. Lett.}\ }\textbf {\bibinfo {volume} {38}},\
  \bibinfo {pages} {380} (\bibinfo {year} {2013})}\BibitemShut {NoStop}%
\bibitem [{\citenamefont {Driben}\ \emph {et~al.}(2013)\citenamefont {Driben},
  \citenamefont {Hu}, \citenamefont {Chen}, \citenamefont {Malomed},\ and\
  \citenamefont {Morandotti}}]{driben.ol.38.2499.2013}%
  \BibitemOpen
  \bibfield  {author} {\bibinfo {author} {\bibfnamefont {R.}~\bibnamefont
  {Driben}}, \bibinfo {author} {\bibfnamefont {Y.}~\bibnamefont {Hu}}, \bibinfo
  {author} {\bibfnamefont {Z.}~\bibnamefont {Chen}}, \bibinfo {author}
  {\bibfnamefont {B.~A.}\ \bibnamefont {Malomed}}, \ and\ \bibinfo {author}
  {\bibfnamefont {R.}~\bibnamefont {Morandotti}},\ }\emph {\bibinfo {title}
  {Inversion and tight focusing of {A}iry pulses under the action of
  third-order dispersion}},\ \href {\doibase 10.1364/OL.38.002499} {\bibfield
  {journal} {\bibinfo  {journal} {Opt. Lett.}\ }\textbf {\bibinfo {volume}
  {38}},\ \bibinfo {pages} {2499} (\bibinfo {year} {2013})}\BibitemShut
  {NoStop}%
\bibitem [{\citenamefont {Zhang}\ \emph
  {et~al.}(2015{\natexlab{a}})\citenamefont {Zhang}, \citenamefont {Liu},
  \citenamefont {Zhong}, \citenamefont {Zhang}, \citenamefont {Li},\ and\
  \citenamefont {Fan}}]{zhang.oe.23.2566.2015}%
  \BibitemOpen
  \bibfield  {author} {\bibinfo {author} {\bibfnamefont {L.}~\bibnamefont
  {Zhang}}, \bibinfo {author} {\bibfnamefont {K.}~\bibnamefont {Liu}}, \bibinfo
  {author} {\bibfnamefont {H.}~\bibnamefont {Zhong}}, \bibinfo {author}
  {\bibfnamefont {J.}~\bibnamefont {Zhang}}, \bibinfo {author} {\bibfnamefont
  {Y.}~\bibnamefont {Li}}, \ and\ \bibinfo {author} {\bibfnamefont
  {D.}~\bibnamefont {Fan}},\ }\emph {\bibinfo {title} {Effect of initial
  frequency chirp on {A}iry pulse propagation in an optical fiber}},\ \href
  {\doibase 10.1364/OE.23.002566} {\bibfield  {journal} {\bibinfo  {journal}
  {Opt. Express}\ }\textbf {\bibinfo {volume} {23}},\ \bibinfo {pages} {2566}
  (\bibinfo {year} {2015}{\natexlab{a}})}\BibitemShut {NoStop}%
\bibitem [{\citenamefont {Hu}\ \emph {et~al.}(2015)\citenamefont {Hu},
  \citenamefont {Tehranchi}, \citenamefont {Wabnitz}, \citenamefont {Kashyap},
  \citenamefont {Chen},\ and\ \citenamefont
  {Morandotti}}]{hu.prl.114.073901.2015}%
  \BibitemOpen
  \bibfield  {author} {\bibinfo {author} {\bibfnamefont {Y.}~\bibnamefont
  {Hu}}, \bibinfo {author} {\bibfnamefont {A.}~\bibnamefont {Tehranchi}},
  \bibinfo {author} {\bibfnamefont {S.}~\bibnamefont {Wabnitz}}, \bibinfo
  {author} {\bibfnamefont {R.}~\bibnamefont {Kashyap}}, \bibinfo {author}
  {\bibfnamefont {Z.}~\bibnamefont {Chen}}, \ and\ \bibinfo {author}
  {\bibfnamefont {R.}~\bibnamefont {Morandotti}},\ }\emph {\bibinfo {title}
  {Improved Intrapulse {R}aman Scattering Control via Asymmetric {A}iry
  Pulses}},\ \href {\doibase 10.1103/PhysRevLett.114.073901} {\bibfield
  {journal} {\bibinfo  {journal} {Phys. Rev. Lett.}\ }\textbf {\bibinfo
  {volume} {114}},\ \bibinfo {pages} {073901} (\bibinfo {year}
  {2015})}\BibitemShut {NoStop}%
\bibitem [{\citenamefont {Bandres}\ and\ \citenamefont
  {Guti\'{e}rrez-Vega}(2007)}]{bandres.oe.15.16719.2007}%
  \BibitemOpen
  \bibfield  {author} {\bibinfo {author} {\bibfnamefont {M.~A.}\ \bibnamefont
  {Bandres}}\ and\ \bibinfo {author} {\bibfnamefont {J.~C.}\ \bibnamefont
  {Guti\'{e}rrez-Vega}},\ }\emph {\bibinfo {title} {{Airy-Gauss} beams and
  their transformation by paraxial optical systems}},\ \href {\doibase
  10.1364/OE.15.016719} {\bibfield  {journal} {\bibinfo  {journal} {Opt.
  Express}\ }\textbf {\bibinfo {volume} {15}},\ \bibinfo {pages} {16719}
  (\bibinfo {year} {2007})}\BibitemShut {NoStop}%
\bibitem [{\citenamefont {Zhang}\ \emph
  {et~al.}(2015{\natexlab{b}})\citenamefont {Zhang}, \citenamefont {Beli\'c},
  \citenamefont {Zhang}, \citenamefont {Zhong}, \citenamefont {Zhu},
  \citenamefont {Wang},\ and\ \citenamefont {Zhang}}]{zhang.oe.23.10467.2015}%
  \BibitemOpen
  \bibfield  {author} {\bibinfo {author} {\bibfnamefont {Y.~Q.}\ \bibnamefont
  {Zhang}}, \bibinfo {author} {\bibfnamefont {M.~R.}\ \bibnamefont {Beli\'c}},
  \bibinfo {author} {\bibfnamefont {L.}~\bibnamefont {Zhang}}, \bibinfo
  {author} {\bibfnamefont {W.~P.}\ \bibnamefont {Zhong}}, \bibinfo {author}
  {\bibfnamefont {D.~Y.}\ \bibnamefont {Zhu}}, \bibinfo {author} {\bibfnamefont
  {R.~M.}\ \bibnamefont {Wang}}, \ and\ \bibinfo {author} {\bibfnamefont
  {Y.~P.}\ \bibnamefont {Zhang}},\ }\emph {\bibinfo {title} {Periodic inversion
  and phase transition of finite energy {A}iry beams in a medium with parabolic
  potential}},\ \href {\doibase 10.1364/OE.23.010467} {\bibfield  {journal}
  {\bibinfo  {journal} {Opt. Express}\ }\textbf {\bibinfo {volume} {23}},\
  \bibinfo {pages} {10467} (\bibinfo {year} {2015}{\natexlab{b}})}\BibitemShut
  {NoStop}%
\bibitem [{\citenamefont {Zhang}\ \emph
  {et~al.}(2015{\natexlab{c}})\citenamefont {Zhang}, \citenamefont {Liu},
  \citenamefont {Beli\'c}, \citenamefont {Zhong}, \citenamefont {Petrovi\'c},\
  and\ \citenamefont {Zhang}}]{zhang.aop.363.305.2015}%
  \BibitemOpen
  \bibfield  {author} {\bibinfo {author} {\bibfnamefont {Y.~Q.}\ \bibnamefont
  {Zhang}}, \bibinfo {author} {\bibfnamefont {X.}~\bibnamefont {Liu}}, \bibinfo
  {author} {\bibfnamefont {M.~R.}\ \bibnamefont {Beli\'c}}, \bibinfo {author}
  {\bibfnamefont {W.~P.}\ \bibnamefont {Zhong}}, \bibinfo {author}
  {\bibfnamefont {M.~S.}\ \bibnamefont {Petrovi\'c}}, \ and\ \bibinfo {author}
  {\bibfnamefont {Y.~P.}\ \bibnamefont {Zhang}},\ }\emph {\bibinfo {title}
  {Automatic {F}ourier transform and self-{F}ourier beams due to parabolic
  potential}},\ \href {\doibase 10.1016/j.aop.2015.10.006} {\bibfield
  {journal} {\bibinfo  {journal} {Ann. Phys.}\ }\textbf {\bibinfo {volume}
  {363}},\ \bibinfo {pages} {305} (\bibinfo {year}
  {2015}{\natexlab{c}})}\BibitemShut {NoStop}%
\bibitem [{\citenamefont {Liu}\ \emph {et~al.}(2011)\citenamefont {Liu},
  \citenamefont {Neshev}, \citenamefont {Shadrivov}, \citenamefont
  {Miroshnichenko},\ and\ \citenamefont {Kivshar}}]{liu.ol.36.1164.2011}%
  \BibitemOpen
  \bibfield  {author} {\bibinfo {author} {\bibfnamefont {W.}~\bibnamefont
  {Liu}}, \bibinfo {author} {\bibfnamefont {D.~N.}\ \bibnamefont {Neshev}},
  \bibinfo {author} {\bibfnamefont {I.~V.}\ \bibnamefont {Shadrivov}}, \bibinfo
  {author} {\bibfnamefont {A.~E.}\ \bibnamefont {Miroshnichenko}}, \ and\
  \bibinfo {author} {\bibfnamefont {Y.~S.}\ \bibnamefont {Kivshar}},\ }\emph
  {\bibinfo {title} {Plasmonic {A}iry beam manipulation in linear optical
  potentials}},\ \href {\doibase 10.1364/OL.36.001164} {\bibfield  {journal}
  {\bibinfo  {journal} {Opt. Lett.}\ }\textbf {\bibinfo {volume} {36}},\
  \bibinfo {pages} {1164} (\bibinfo {year} {2011})}\BibitemShut {NoStop}%
\bibitem [{\citenamefont {Efremidis}(2011)}]{efremidis.ol.36.3006.2011}%
  \BibitemOpen
  \bibfield  {author} {\bibinfo {author} {\bibfnamefont {N.~K.}\ \bibnamefont
  {Efremidis}},\ }\emph {\bibinfo {title} {Airy trajectory engineering in
  dynamic linear index potentials}},\ \href {\doibase 10.1364/OL.36.003006}
  {\bibfield  {journal} {\bibinfo  {journal} {Opt. Lett.}\ }\textbf {\bibinfo
  {volume} {36}},\ \bibinfo {pages} {3006} (\bibinfo {year}
  {2011})}\BibitemShut {NoStop}%
\bibitem [{\citenamefont {Hu}\ \emph {et~al.}(2012)\citenamefont {Hu},
  \citenamefont {Siviloglou}, \citenamefont {Zhang}, \citenamefont {Efremidis},
  \citenamefont {Christodoulides},\ and\ \citenamefont {Chen}}]{hu.book.2012}%
  \BibitemOpen
  \bibfield  {author} {\bibinfo {author} {\bibfnamefont {Y.}~\bibnamefont
  {Hu}}, \bibinfo {author} {\bibfnamefont {G.~A.}\ \bibnamefont {Siviloglou}},
  \bibinfo {author} {\bibfnamefont {P.}~\bibnamefont {Zhang}}, \bibinfo
  {author} {\bibfnamefont {N.~K.}\ \bibnamefont {Efremidis}}, \bibinfo {author}
  {\bibfnamefont {D.~N.}\ \bibnamefont {Christodoulides}}, \ and\ \bibinfo
  {author} {\bibfnamefont {Z.}~\bibnamefont {Chen}},\ }in\ \href {\doibase
  10.1007/978-1-4614-3538-9_1} {\emph {\bibinfo {booktitle} {Nonlinear
  Photonics and Novel Optical Phenomena}}},\ \bibinfo {series} {Springer Series
  in Optical Sciences}, Vol.\ \bibinfo {volume} {170},\ \bibinfo {editor}
  {edited by\ \bibinfo {editor} {\bibfnamefont {Z.}~\bibnamefont {Chen}}\ and\
  \bibinfo {editor} {\bibfnamefont {R.}~\bibnamefont {Morandotti}}}\ (\bibinfo
  {publisher} {Springer},\ \bibinfo {address} {New York},\ \bibinfo {year}
  {2012})\ pp.\ \bibinfo {pages} {1--46}\BibitemShut {NoStop}%
\bibitem [{\citenamefont {Zhang}\ \emph
  {et~al.}(2013{\natexlab{b}})\citenamefont {Zhang}, \citenamefont {Hu},
  \citenamefont {Zhao}, \citenamefont {Zhang},\ and\ \citenamefont
  {Chen}}]{zhang.csb.58.3513.2013}%
  \BibitemOpen
  \bibfield  {author} {\bibinfo {author} {\bibfnamefont {Z.}~\bibnamefont
  {Zhang}}, \bibinfo {author} {\bibfnamefont {Y.}~\bibnamefont {Hu}}, \bibinfo
  {author} {\bibfnamefont {J.~Y.}\ \bibnamefont {Zhao}}, \bibinfo {author}
  {\bibfnamefont {P.}~\bibnamefont {Zhang}}, \ and\ \bibinfo {author}
  {\bibfnamefont {Z.~G.}\ \bibnamefont {Chen}},\ }\emph {\bibinfo {title}
  {Research progress and application prospect of {A}iry beams}},\ \href
  {\doibase 10.1360/972013-1125} {\bibfield  {journal} {\bibinfo  {journal}
  {Chin. Sci. Bull.}\ }\textbf {\bibinfo {volume} {58}},\ \bibinfo {pages}
  {3513} (\bibinfo {year} {2013}{\natexlab{b}})}\BibitemShut {NoStop}%
\bibitem [{\citenamefont {Bandres}\ \emph {et~al.}(2013)\citenamefont
  {Bandres}, \citenamefont {Kaminer}, \citenamefont {Mills}, \citenamefont
  {Rodriguez-Lara}, \citenamefont {Greenfield}, \citenamefont {Segev},\ and\
  \citenamefont {Christodoulides}}]{bandres.opn.24.30.2013}%
  \BibitemOpen
  \bibfield  {author} {\bibinfo {author} {\bibfnamefont {M.~A.}\ \bibnamefont
  {Bandres}}, \bibinfo {author} {\bibfnamefont {I.}~\bibnamefont {Kaminer}},
  \bibinfo {author} {\bibfnamefont {M.}~\bibnamefont {Mills}}, \bibinfo
  {author} {\bibfnamefont {B.~M.}\ \bibnamefont {Rodriguez-Lara}}, \bibinfo
  {author} {\bibfnamefont {E.}~\bibnamefont {Greenfield}}, \bibinfo {author}
  {\bibfnamefont {M.}~\bibnamefont {Segev}}, \ and\ \bibinfo {author}
  {\bibfnamefont {D.~N.}\ \bibnamefont {Christodoulides}},\ }\emph {\bibinfo
  {title} {Accelerating Optical Beams}},\ \href {\doibase
  10.1364/OPN.24.6.000030} {\bibfield  {journal} {\bibinfo  {journal} {Opt.
  Phot. News}\ }\textbf {\bibinfo {volume} {24}},\ \bibinfo {pages} {30}
  (\bibinfo {year} {2013})}\BibitemShut {NoStop}%
\bibitem [{\citenamefont {Efremidis}\ and\ \citenamefont
  {Christodoulides}(2010)}]{efremidis.ol.35.4045.2010}%
  \BibitemOpen
  \bibfield  {author} {\bibinfo {author} {\bibfnamefont {N.~K.}\ \bibnamefont
  {Efremidis}}\ and\ \bibinfo {author} {\bibfnamefont {D.~N.}\ \bibnamefont
  {Christodoulides}},\ }\emph {\bibinfo {title} {Abruptly autofocusing
  waves}},\ \href {\doibase 10.1364/OL.35.004045} {\bibfield  {journal}
  {\bibinfo  {journal} {Opt. Lett.}\ }\textbf {\bibinfo {volume} {35}},\
  \bibinfo {pages} {4045} (\bibinfo {year} {2010})}\BibitemShut {NoStop}%
\bibitem [{\citenamefont {Papazoglou}\ \emph {et~al.}(2011)\citenamefont
  {Papazoglou}, \citenamefont {Efremidis}, \citenamefont {Christodoulides},\
  and\ \citenamefont {Tzortzakis}}]{papazoglou.ol.36.1842.2011}%
  \BibitemOpen
  \bibfield  {author} {\bibinfo {author} {\bibfnamefont {D.~G.}\ \bibnamefont
  {Papazoglou}}, \bibinfo {author} {\bibfnamefont {N.~K.}\ \bibnamefont
  {Efremidis}}, \bibinfo {author} {\bibfnamefont {D.~N.}\ \bibnamefont
  {Christodoulides}}, \ and\ \bibinfo {author} {\bibfnamefont {S.}~\bibnamefont
  {Tzortzakis}},\ }\emph {\bibinfo {title} {Observation of abruptly
  autofocusing waves}},\ \href {\doibase 10.1364/OL.36.001842} {\bibfield
  {journal} {\bibinfo  {journal} {Opt. Lett.}\ }\textbf {\bibinfo {volume}
  {36}},\ \bibinfo {pages} {1842} (\bibinfo {year} {2011})}\BibitemShut
  {NoStop}%
\bibitem [{\citenamefont {Chremmos}\ \emph
  {et~al.}(2011{\natexlab{a}})\citenamefont {Chremmos}, \citenamefont
  {Efremidis},\ and\ \citenamefont
  {Christodoulides}}]{chremmos.ol.36.1890.2011}%
  \BibitemOpen
  \bibfield  {author} {\bibinfo {author} {\bibfnamefont {I.}~\bibnamefont
  {Chremmos}}, \bibinfo {author} {\bibfnamefont {N.~K.}\ \bibnamefont
  {Efremidis}}, \ and\ \bibinfo {author} {\bibfnamefont {D.~N.}\ \bibnamefont
  {Christodoulides}},\ }\emph {\bibinfo {title} {Pre-engineered abruptly
  autofocusing beams}},\ \href {\doibase 10.1364/OL.36.001890} {\bibfield
  {journal} {\bibinfo  {journal} {Opt. Lett.}\ }\textbf {\bibinfo {volume}
  {36}},\ \bibinfo {pages} {1890} (\bibinfo {year}
  {2011}{\natexlab{a}})}\BibitemShut {NoStop}%
\bibitem [{\citenamefont {Penciu}\ \emph {et~al.}(2016)\citenamefont {Penciu},
  \citenamefont {Makris},\ and\ \citenamefont
  {Efremidis}}]{penciu.ol.41.1042.2016}%
  \BibitemOpen
  \bibfield  {author} {\bibinfo {author} {\bibfnamefont {R.-S.}\ \bibnamefont
  {Penciu}}, \bibinfo {author} {\bibfnamefont {K.~G.}\ \bibnamefont {Makris}},
  \ and\ \bibinfo {author} {\bibfnamefont {N.~K.}\ \bibnamefont {Efremidis}},\
  }\emph {\bibinfo {title} {Nonparaxial abruptly autofocusing beams}},\ \href
  {\doibase 10.1364/OL.41.001042} {\bibfield  {journal} {\bibinfo  {journal}
  {Opt. Lett.}\ }\textbf {\bibinfo {volume} {41}},\ \bibinfo {pages} {1042}
  (\bibinfo {year} {2016})}\BibitemShut {NoStop}%
\bibitem [{\citenamefont {Zhang}\ \emph
  {et~al.}(2011{\natexlab{b}})\citenamefont {Zhang}, \citenamefont {Prakash},
  \citenamefont {Zhang}, \citenamefont {Mills}, \citenamefont {Efremidis},
  \citenamefont {Christodoulides},\ and\ \citenamefont
  {Chen}}]{zhang.ol.36.2883.2011}%
  \BibitemOpen
  \bibfield  {author} {\bibinfo {author} {\bibfnamefont {P.}~\bibnamefont
  {Zhang}}, \bibinfo {author} {\bibfnamefont {J.}~\bibnamefont {Prakash}},
  \bibinfo {author} {\bibfnamefont {Z.}~\bibnamefont {Zhang}}, \bibinfo
  {author} {\bibfnamefont {M.~S.}\ \bibnamefont {Mills}}, \bibinfo {author}
  {\bibfnamefont {N.~K.}\ \bibnamefont {Efremidis}}, \bibinfo {author}
  {\bibfnamefont {D.~N.}\ \bibnamefont {Christodoulides}}, \ and\ \bibinfo
  {author} {\bibfnamefont {Z.}~\bibnamefont {Chen}},\ }\emph {\bibinfo {title}
  {Trapping and guiding microparticles with morphing autofocusing {A}iry
  beams}},\ \href {\doibase 10.1364/OL.36.002883} {\bibfield  {journal}
  {\bibinfo  {journal} {Opt. Lett.}\ }\textbf {\bibinfo {volume} {36}},\
  \bibinfo {pages} {2883} (\bibinfo {year} {2011}{\natexlab{b}})}\BibitemShut
  {NoStop}%
\bibitem [{\citenamefont {Li}\ \emph {et~al.}(2014)\citenamefont {Li},
  \citenamefont {Liu}, \citenamefont {Peng}, \citenamefont {Xie}, \citenamefont
  {Gan},\ and\ \citenamefont {Zhao}}]{li.oe.22.7598.2014}%
  \BibitemOpen
  \bibfield  {author} {\bibinfo {author} {\bibfnamefont {P.}~\bibnamefont
  {Li}}, \bibinfo {author} {\bibfnamefont {S.}~\bibnamefont {Liu}}, \bibinfo
  {author} {\bibfnamefont {T.}~\bibnamefont {Peng}}, \bibinfo {author}
  {\bibfnamefont {G.}~\bibnamefont {Xie}}, \bibinfo {author} {\bibfnamefont
  {X.}~\bibnamefont {Gan}}, \ and\ \bibinfo {author} {\bibfnamefont
  {J.}~\bibnamefont {Zhao}},\ }\emph {\bibinfo {title} {Spiral autofocusing
  {A}iry beams carrying power-exponent-phase vortices}},\ \href {\doibase
  10.1364/OE.22.007598} {\bibfield  {journal} {\bibinfo  {journal} {Opt.
  Express}\ }\textbf {\bibinfo {volume} {22}},\ \bibinfo {pages} {7598}
  (\bibinfo {year} {2014})}\BibitemShut {NoStop}%
\bibitem [{\citenamefont {Chremmos}\ \emph
  {et~al.}(2011{\natexlab{b}})\citenamefont {Chremmos}, \citenamefont {Zhang},
  \citenamefont {Prakash}, \citenamefont {Efremidis}, \citenamefont
  {Christodoulides},\ and\ \citenamefont {Chen}}]{chremmos.ol.36.3675.2011}%
  \BibitemOpen
  \bibfield  {author} {\bibinfo {author} {\bibfnamefont {I.}~\bibnamefont
  {Chremmos}}, \bibinfo {author} {\bibfnamefont {P.}~\bibnamefont {Zhang}},
  \bibinfo {author} {\bibfnamefont {J.}~\bibnamefont {Prakash}}, \bibinfo
  {author} {\bibfnamefont {N.~K.}\ \bibnamefont {Efremidis}}, \bibinfo {author}
  {\bibfnamefont {D.~N.}\ \bibnamefont {Christodoulides}}, \ and\ \bibinfo
  {author} {\bibfnamefont {Z.}~\bibnamefont {Chen}},\ }\emph {\bibinfo {title}
  {Fourier-space generation of abruptly autofocusing beams and optical bottle
  beams}},\ \href {\doibase 10.1364/OL.36.003675} {\bibfield  {journal}
  {\bibinfo  {journal} {Opt. Lett.}\ }\textbf {\bibinfo {volume} {36}},\
  \bibinfo {pages} {3675} (\bibinfo {year} {2011}{\natexlab{b}})}\BibitemShut
  {NoStop}%
\bibitem [{\citenamefont {Panagiotopoulos}\ \emph {et~al.}(2013)\citenamefont
  {Panagiotopoulos}, \citenamefont {Papazoglou}, \citenamefont {Couairon},\
  and\ \citenamefont {Tzortzakis}}]{Panagiotopoulos.nc.4.2622.2013}%
  \BibitemOpen
  \bibfield  {author} {\bibinfo {author} {\bibfnamefont {P.}~\bibnamefont
  {Panagiotopoulos}}, \bibinfo {author} {\bibfnamefont {D.~G.}\ \bibnamefont
  {Papazoglou}}, \bibinfo {author} {\bibfnamefont {A.}~\bibnamefont
  {Couairon}}, \ and\ \bibinfo {author} {\bibfnamefont {S.}~\bibnamefont
  {Tzortzakis}},\ }\emph {\bibinfo {title} {Sharply autofocused ring-{A}iry
  beams transforming into non-linear intense light bullets}},\ \href {\doibase
  10.1038/ncomms3622} {\bibfield  {journal} {\bibinfo  {journal} {Nat.
  Commun.}\ }\textbf {\bibinfo {volume} {4}},\ \bibinfo {pages} {2622}
  (\bibinfo {year} {2013})}\BibitemShut {NoStop}%
\bibitem [{\citenamefont {Hwang}\ \emph {et~al.}(2012)\citenamefont {Hwang},
  \citenamefont {Kim},\ and\ \citenamefont {Lee}}]{hwang.pj.4.174.2012}%
  \BibitemOpen
  \bibfield  {author} {\bibinfo {author} {\bibfnamefont {C.-Y.}\ \bibnamefont
  {Hwang}}, \bibinfo {author} {\bibfnamefont {K.~Y.}\ \bibnamefont {Kim}}, \
  and\ \bibinfo {author} {\bibfnamefont {B.}~\bibnamefont {Lee}},\ }\emph
  {\bibinfo {title} {Dynamic Control of Circular {A}iry Beams With Linear
  Optical Potentials}},\ \href {\doibase 10.1109/JPHOT.2011.2182338} {\bibfield
   {journal} {\bibinfo  {journal} {IEEE Photon. J.}\ }\textbf {\bibinfo
  {volume} {4}},\ \bibinfo {pages} {174} (\bibinfo {year} {2012})}\BibitemShut
  {NoStop}%
\bibitem [{\citenamefont {Chremmos}\ \emph {et~al.}(2012)\citenamefont
  {Chremmos}, \citenamefont {Chen}, \citenamefont {Christodoulides},\ and\
  \citenamefont {Efremidis}}]{ioannis.pra.85.023828.2012}%
  \BibitemOpen
  \bibfield  {author} {\bibinfo {author} {\bibfnamefont {I.~D.}\ \bibnamefont
  {Chremmos}}, \bibinfo {author} {\bibfnamefont {Z.}~\bibnamefont {Chen}},
  \bibinfo {author} {\bibfnamefont {D.~N.}\ \bibnamefont {Christodoulides}}, \
  and\ \bibinfo {author} {\bibfnamefont {N.~K.}\ \bibnamefont {Efremidis}},\
  }\emph {\bibinfo {title} {Abruptly autofocusing and autodefocusing optical
  beams with arbitrary caustics}},\ \href {\doibase 10.1103/PhysRevA.85.023828}
  {\bibfield  {journal} {\bibinfo  {journal} {Phys. Rev. A}\ }\textbf {\bibinfo
  {volume} {85}},\ \bibinfo {pages} {023828} (\bibinfo {year}
  {2012})}\BibitemShut {NoStop}%
\bibitem [{\citenamefont {Zhang}\ \emph
  {et~al.}(2015{\natexlab{d}})\citenamefont {Zhang}, \citenamefont {Liu},
  \citenamefont {Beli\'c}, \citenamefont {Zhong}, \citenamefont {Wen},\ and\
  \citenamefont {Zhang}}]{zhang.ol.40.3786.2015}%
  \BibitemOpen
  \bibfield  {author} {\bibinfo {author} {\bibfnamefont {Y.~Q.}\ \bibnamefont
  {Zhang}}, \bibinfo {author} {\bibfnamefont {X.}~\bibnamefont {Liu}}, \bibinfo
  {author} {\bibfnamefont {M.~R.}\ \bibnamefont {Beli\'c}}, \bibinfo {author}
  {\bibfnamefont {W.~P.}\ \bibnamefont {Zhong}}, \bibinfo {author}
  {\bibfnamefont {F.}~\bibnamefont {Wen}}, \ and\ \bibinfo {author}
  {\bibfnamefont {Y.~P.}\ \bibnamefont {Zhang}},\ }\emph {\bibinfo {title}
  {Anharmonic propagation of two-dimensional beams carrying orbital angular
  momentum in a harmonic potential}},\ \href {\doibase 10.1364/OL.40.003786}
  {\bibfield  {journal} {\bibinfo  {journal} {Opt. Lett.}\ }\textbf {\bibinfo
  {volume} {40}},\ \bibinfo {pages} {3786} (\bibinfo {year}
  {2015}{\natexlab{d}})}\BibitemShut {NoStop}%
\bibitem [{\citenamefont {Zhang}\ \emph
  {et~al.}(2015{\natexlab{e}})\citenamefont {Zhang}, \citenamefont {Liu},
  \citenamefont {Beli\'{c}}, \citenamefont {Zhong}, \citenamefont {Zhang},\
  and\ \citenamefont {Xiao}}]{zhang.prl.115.180403.2015}%
  \BibitemOpen
  \bibfield  {author} {\bibinfo {author} {\bibfnamefont {Y.~Q.}\ \bibnamefont
  {Zhang}}, \bibinfo {author} {\bibfnamefont {X.}~\bibnamefont {Liu}}, \bibinfo
  {author} {\bibfnamefont {M.~R.}\ \bibnamefont {Beli\'{c}}}, \bibinfo {author}
  {\bibfnamefont {W.~P.}\ \bibnamefont {Zhong}}, \bibinfo {author}
  {\bibfnamefont {Y.~P.}\ \bibnamefont {Zhang}}, \ and\ \bibinfo {author}
  {\bibfnamefont {M.}~\bibnamefont {Xiao}},\ }\emph {\bibinfo {title}
  {Propagation Dynamics of a Light Beam in a Fractional {S}chr\"odinger
  Equation}},\ \href {\doibase 10.1103/PhysRevLett.115.180403} {\bibfield
  {journal} {\bibinfo  {journal} {Phys. Rev. Lett.}\ }\textbf {\bibinfo
  {volume} {115}},\ \bibinfo {pages} {180403} (\bibinfo {year}
  {2015}{\natexlab{e}})}\BibitemShut {NoStop}%
\bibitem [{\citenamefont {Zhang}\ \emph
  {et~al.}(2013{\natexlab{c}})\citenamefont {Zhang}, \citenamefont {Liu},
  \citenamefont {Zhang}, \citenamefont {Ni}, \citenamefont {Prakash},
  \citenamefont {Hu}, \citenamefont {Jiang}, \citenamefont {Christodoulides},\
  and\ \citenamefont {Chen}}]{zhang.wulixuebao.62.34209.2013}%
  \BibitemOpen
  \bibfield  {author} {\bibinfo {author} {\bibfnamefont {Z.}~\bibnamefont
  {Zhang}}, \bibinfo {author} {\bibfnamefont {J.-J.}\ \bibnamefont {Liu}},
  \bibinfo {author} {\bibfnamefont {P.}~\bibnamefont {Zhang}}, \bibinfo
  {author} {\bibfnamefont {P.-G.}\ \bibnamefont {Ni}}, \bibinfo {author}
  {\bibfnamefont {J.}~\bibnamefont {Prakash}}, \bibinfo {author} {\bibfnamefont
  {Y.}~\bibnamefont {Hu}}, \bibinfo {author} {\bibfnamefont {D.-S.}\
  \bibnamefont {Jiang}}, \bibinfo {author} {\bibfnamefont {D.~N.}\ \bibnamefont
  {Christodoulides}}, \ and\ \bibinfo {author} {\bibfnamefont {Z.-G.}\
  \bibnamefont {Chen}},\ }\emph {\bibinfo {title} {Generation of autofocusing
  beams with multi-{A}iry beams}},\ \href {\doibase 10.7498/aps.62.034209}
  {\bibfield  {journal} {\bibinfo  {journal} {Acta Phys. Sin.}\ }\textbf
  {\bibinfo {volume} {62}},\ \bibinfo {pages} {034209} (\bibinfo {year}
  {2013}{\natexlab{c}})}\BibitemShut {NoStop}%
\bibitem [{\citenamefont {Hwang}\ \emph {et~al.}(2011)\citenamefont {Hwang},
  \citenamefont {Kim},\ and\ \citenamefont {Lee}}]{hwang.oe.19.7356.2011}%
  \BibitemOpen
  \bibfield  {author} {\bibinfo {author} {\bibfnamefont {C.-Y.}\ \bibnamefont
  {Hwang}}, \bibinfo {author} {\bibfnamefont {K.-Y.}\ \bibnamefont {Kim}}, \
  and\ \bibinfo {author} {\bibfnamefont {B.}~\bibnamefont {Lee}},\ }\emph
  {\bibinfo {title} {Bessel-like beam generation by superposing multiple {A}iry
  beams}},\ \href {\doibase 10.1364/OE.19.007356} {\bibfield  {journal}
  {\bibinfo  {journal} {Opt. Express}\ }\textbf {\bibinfo {volume} {19}},\
  \bibinfo {pages} {7356} (\bibinfo {year} {2011})}\BibitemShut {NoStop}%
\end{thebibliography}%

\end{document}